\documentclass[aps,twocolumn,prc,superscriptaddress,showpacs,nofootinbib,floatfix,amssymb,amsfonts,amsmath]{revtex4-1}
\usepackage{graphicx}
\usepackage{dcolumn}
\usepackage{bm}
\usepackage{amsmath}    
\usepackage{amsfonts}   
\usepackage{amssymb}
\usepackage{graphicx}   

\begin{document}
\title{Assessing theoretical uncertainties in fission barriers of
superheavy nuclei.}

\author{S.\ E.\ Agbemava}
\affiliation{Department of Physics and Astronomy, Mississippi
State University, MS 39762}

\author{A.\ V.\ Afanasjev}
\affiliation{Department of Physics and Astronomy, Mississippi
State University, MS 39762}

\author{D.\ Ray}
\affiliation{Department of Physics and Astronomy, Mississippi
State University, MS 39762}


\author{P. Ring}
\affiliation{Fakult\"at f\"ur Physik, Technische Universit\"at M\"unchen,
 D-85748 Garching, Germany}

\date{\today}

\begin{abstract}
   Theoretical uncertainties in the predictions of inner fission barrier
heights in superheavy elements have been investigated in a systematic
way for a set of state-of-the-art covariant energy density functionals
which represent major classes of the functionals used in covariant
density functional theory. They differ in basic model assumptions and
fitting protocols. Both systematic and statistical uncertainties have
been quantified where the former turn out to be larger. Systematic uncertainties
are substantial in superheavy elements and their behavior as a function
of proton and neutron numbers contains a large random component. The
benchmarking of the functionals to the experimental data on fission
barriers in the actinides allows to reduce the systematic theoretical
uncertainties for the inner fission barriers of unknown superheavy
elements. However, even then they on average increase on moving away
from the region where benchmarking has been performed. In addition,
a comparison with the results of non-relativistic approaches is
performed in order to define full systematic theoretical uncertainties
over the state-of-the-art models. Even for the models benchmarked in
the actinides, the difference in the inner fission barrier height
of some superheavy elements reaches $5-6$ MeV. This uncertainty in
the fission barrier heights will translate into huge (many tens of the
orders of magnitude) uncertainties in the spontaneous fission half-lives.
\end{abstract}


\maketitle

\section{Introduction}

  The region of superheavy elements (SHE), characterized by the
extreme values of proton number $Z$, is one of the extremes of the
nuclear landscape and an arena of active experimental and
theoretical studies (see Refs.\ \cite{SP.07,OU.15,AANR.15} and
references therein). Contrary to other regions of the nuclear chart,
the SHE are stabilized only by quantum shell effects. Currently
available experimental data reach proton number $Z=118$
\cite{Z=116-118-year2006,Z=117-118-year2012} and dedicated
experimental facilities such as the Dubna Superheavy Element Factory
will hopefully allow to extend the region of SHE up to $Z=120$ and
for a wider range of neutron numbers for lower $Z$ values.

  The stability of SHEs is defined by the fission barriers. In addition,
the experimental studies of SHEs are based on the observation of
$\alpha$-decays. As a consequence, only SHEs with spontaneous fission
half-lives $\tau_{SF}$ longer than the half-lives $\tau_{\alpha}$ of the
$\alpha$-decays could be observed in experiment. An additional limit is set
up by the fact that only $\alpha$-decays longer than 10 $\mu$s can be
observed in experiment. Therefore it is of great importance to study the fission
barriers in SHEs. The height of the fission barrier, $B_f$, which is
the difference of the energies of the respective saddle in the potential
energy surface (PES) and the ground  state, is one of most important
quantities. It defines the survival probability of SHEs synthesized in
heavy-ion reactions and impacts the spontaneous fission half-lives. The
later is important for an understanding of the competition between the
fission process and $\alpha$ particle emission.

   Fission barriers have been extensively studied in different theoretical
frameworks; these studies have been reviewed in Refs.\ \cite{SP.07,BKRRSW.15}.
The theoretical frameworks used are the microscopic+macroscopic method
\cite{NilRag-book}, non-relativistic density functional theories (DFT) based
on finite range Gogny \cite{PM.14} and zero range Skyrme forces \cite{BHP.03},
and covariant density functional theory (CDFT) \cite{VALR.05}. Our present
investigation is performed in CDFT. It has been less frequently used in the
studies of fission barriers in SHEs as compared with non-relativistic theories:
a systematic investigation of the fission barriers in
the $Z=112-120$ SHE has been performed in the triaxial relativistic mean field plus
BCS (RMF+BCS) framework with the NL3* functional in Ref.\ \cite{AAR.12} and
potential energy surfaces in the $(\beta,\gamma)$ plane for the even-even isotopes
in the $\alpha$-decay chains of the $^{298}120$ and
$^{300}120$ nuclei have been calculated in the triaxial relativistic Hartree-Bogoliubov
approach with the DD-PC1 functional in Ref.\ \cite{PNLV.12}.

Theoretical investigations require an estimate of theoretical uncertainties.
This becomes especially important when one deals with the extrapolations beyond
the known regions, as for example in particle number or deformation. This
issue has been discussed in detail in Refs.\ \cite{RN.10,DNR.14} and in
the context of global studies  within CDFT in the introduction of Ref.\
\cite{AARR.14}. In the CDFT framework, the studies of theoretical uncertainties
have been restricted to the ground state properties so far.  Systematic theoretical
uncertainties and their sources have been studied globally for the ground state masses,
deformations, charge radii, neutrons skins, positions of drip lines etc in Refs.\
\cite{AARR.13,AARR.14,AARR.15,AANR.15,AAR.16,AA.16}. Of particular importance in
the context of the present manuscript is the study of theoretical uncertainties in
the ground state properties of SHE presented in Ref.\ \cite{AANR.15}.  An analysis
of statistical theoretical uncertainties in the ground state observables is currently
underway and will be submitted for publication soon \cite{AA.16-prep}. The major goal of
the present paper is to extend these investigations to excited states, namely,
to the fission barriers in superheavy nuclei. Both statistical and systematic
theoretical uncertainties in the description of fission barriers will be considered
here.

  Theoretical uncertainties emerge from the underlying theoretical approximations.
In the DFT framework, there are two major sources of these approximations, namely,
the range of interaction and the form of the density dependence of the effective interaction
\cite{BHP.03,BB.77}. In the non-relativistic case one has zero range Skyrme and finite range
Gogny forces and different density dependencies \cite{BHP.03}. A similar situation exists
also in the relativistic case: point coupling and meson exchange models have
an interaction of zero and of finite range, respectively \cite{VALR.05,DD-ME2,NL3*,DD-PC1}.
The density dependence is introduced either through an explicit dependence of the coupling
constants \cite{TW.99,DD-ME2,DD-PC1} or via non-linear meson couplings \cite{BB.77,NL3*}.
This ambiguity in the definition of the range of the interaction and its density dependence
leads to several major classes of the covariant energy density functionals (CEDF) which
were discussed in Ref.\ \cite{AARR.14}.

  As a consequence, in the present manuscript, we focus on the uncertainties
related to the choice of the energy density functional. They can be relatively easily
deduced globally \cite{AARR.14} (at least for axial reflection symmetric shapes). We
therefore define theoretical uncertainty for a given physical observable (which we call
in the following ``spreads'')  via the spread of theoretical  predictions as
\cite{AARR.14}
\begin{equation}
\Delta O(Z,N) = |O_{max}(Z,N) - O_{min}(Z,N)|,
\end{equation}
where $O_{max}(Z,N)$ and $O_{min}(Z,N)$ are the largest and smallest
values of the physical observable $O(Z,N)$ obtained within the set
of CEDFs under investigation
for the $(Z,N)$ nucleus. Note that these spreads are only a
crude approximation to the {\it systematic} theoretical errors discussed
in Ref.\ \cite{DNR.14} since they are obtained with a very small number of
functionals which do not form an independent statistical ensemble.
Note also that these {\it systematic} errors are not well defined in unknown
regions of nuclear chart or deformation since systematic biases of
theoretical models could not be established in these regions in the
absence of experimental data and/or an exact theory.


  We use the CEDFs NL3* \cite{NL3*}, DD-ME2 \cite{DD-ME2}, DD-ME$\delta$
\cite{DD-MEdelta}, DD-PC1 \cite{DD-PC1} and PC-PK1 \cite{PC-PK1}.
These state-of-the-art functionals represent the essential types
of CEDFs used in the literature (for more details see the discussion
in Sect. II of Ref.\ \cite{AARR.14} and the introduction to Ref.\
\cite{AANR.15}). Moreover, their performance and the related theoretical
uncertainties have recently been analyzed globally in Refs.\
\cite{AARR.14,ZNLYM.14,AA.16,AAR.16}
and in particular in superheavy nuclei in Ref.\ \cite{AANR.15}. They
are characterized by an improved accuracy of the  description of
experimental data as compared with the previous generation of CEDFs.

In details they are based on rather different concepts:

\begin{itemize}

\item
NL3* \cite{NL3*}, a slightly improved modern version
of the well known functional NL3 \cite{NL3}, is a representative of
the first group of CEDFs proposed in 1977 in
the pioneering work of Boguta and Bodmer \cite{BB.77}.
These two functionals are based
on the Walecka model \cite{Wal.74} with its three mesons $\sigma$,
$\omega$, and $\rho$ and include a density dependence through
non-linear meson couplings in the $\sigma$-channel.
In addition to the four basic parameters ($m_\sigma$, $g_\sigma$, $g_\omega$,
and $g_\rho$) they depend on two non-linear coupling constants $g_2$ and $g_3$
describing the strength of cubic ($\sigma^3$) and quartic ($\sigma^4$) terms.
This class of functionals misses a density dependence in the
isovector channel and therefore the asymmetry energy
of such functionals is relatively large and their dependence on
the density is rather stiff.

\item
The second class of the functionals, originally introduced in 1999 by Typel and Wolter \cite{TW.99}, is also based on meson-exchange forces, but the non-linear meson couplings are replaced by an explicit density dependence of the coupling constants ($g_i(\rho)$, $i=\sigma$, $\omega$, $\rho$) with four additional parameters.
The set DD-ME2 \cite{DD-ME2} is probably one
of the most successful CEDFs of this type. Its eight parameters have
been adjusted in a very careful way to the binding energies and radii of a
set of twelve spherical nuclei. Here $g_\rho(\rho)$ depends on the density
and therefore this set reproduces rather well not only the ab-initio
results for the equation of state (EoS) of symmetric nuclear matter, but also
those for neutron matter \cite{APR.98}.

\item
The third functional DD-ME$\delta$~\cite{DD-MEdelta} is in its form very similar
to DD-ME2 but it represents a new idea. It is, to a large extent, derived from
modern ab-initio calculations of nuclear matter
\cite{Baldo2008_PLB663-390,VanDalen2007_EPJA31-29}.
Therefore it contains in addition to three mesons $\sigma$,
$\omega$, and $\rho$ the scalar isovector meson $\delta$.
Only four phenomenological parameters ($m_\sigma$, $g_\sigma$, $g_\omega$, and
$g_\rho$) at saturation density are adjusted to the same set of
data as it has been used for DD-ME2. All the rest is derived from ab-initio
calculations.

\item
The last two functionals DD-PC1 \cite{DD-PC1} and PC-PK1 \cite{PC-PK1} have been
chosen because they represent zero-range functionals, which are technically much
simpler than those based on meson exchange forces with finite range. They
can be derived in the limit of large meson masses. This class of
functionals has been first proposed in the eighties by Manakos et al. \cite{MM.88},
but only recently their density dependence has been adjusted carefully to experimental
data. We chose two versions of this model with a different density dependence and
with a different fitting strategy. The functional DD-PC1 \cite{DD-PC1} contains an
exponential density dependence and it has been adjusted only to nuclear matter data
and masses of a large set of deformed nuclei. On the other hand, PC-PK1 \cite{PC-PK1}
has a density dependence of polynomial form in all spin-isospin channels and it is
adjusted to a very large set of spherical nuclei. Because of its polynomial
density dependence it can also be used for beyond mean field calculations
in the framework of the Generator Coordinate Method (GCM)~\cite{YHLMR.14}.
However, it does not have a density dependence in the isovector channel.

\end{itemize}

 An additional source of theoretical uncertainties is related to the
details of the fitting protocol such as the choice of experimental data
and the selection of adopted errors. It applies only to a given functional
and the related theoretical uncertainties are called {\it statistical}
\cite{Stat-an,DNR.14}.
 Note that the selection of adopted errors is to a
degree subjective, in particular, if one deals with quantities of
different dimensions.  The investigation of statistical theoretical uncertainties
for potential energy curves is time-consuming since it involves
constrained deformed calculations over a substantial number of grid
points performed for a substantial number of the variations of the
original functional. As a result, such an analysis is performed only
for a single nucleus and only for two CEDFs.

 We restrict our investigation to inner fission barriers. There are
several reasons behind this choice. A systematic investigation of
Ref.\ \cite{AAR.12} within the RMF+BCS framework with the NL3* CEDF
has shown that the fission barriers of many SHEs have a double-humped
structure in axial reflection-symmetric calculations.
The inclusion of octupole and triaxial deformations lowers
outer fission barriers by 2 to 4 MeV so that they are only around
2 MeV in height with respect of superdeformed minimum. A similar situation
exists also in Gogny DFT calculations (Ref.\ \cite{WE.12}). In addition,
similar to actinides  (Ref.\ \cite{LZZ.14}) symmetry  unrestricted
calculations which combine octupole and triaxial deformations simultaneously
could further reduce the heights of outer fission barriers. These low
barriers would translate into a high penetration probability for spontaneous
fission such that most likely these superdeformed states are metastable and
that outer fission barriers do not affect substantially the fission process
in total.  Note also that outer fission barriers do not exist in most of the
SHEs with $Z\geq110$ in Skyrme DFT calculations \cite{BBM.04,SBN.13}. An accurate
description of outer fission barriers would require the use of a symmetry
unrestricted RHB code. Unfortunately, the
computational cost for such an investigation of theoretical uncertainties
in the description of outer fission barriers is prohibitively high.

  Despite these limitations this investigation provides for a first time
a systematic analysis of theoretical uncertainties in the description of
fission barriers within the CDFT framework. It also gives an
understanding which observables/aspects of many-body physics can be predicted
with a higher level of confidence than others for density functionals of
the given type. Moreover, it is expected that they will indicate which
aspects of many-body problem have to be addressed with more care during
the development of next generation of EDFs. This study also represents
an extension of our previous studies of theoretical uncertainties in the
global description of the ground state properties of the nuclei from the
proton to neutron drip lines \cite{AARR.13,AARR.14,AARR.15,AA.16},
superheavy nuclei \cite{AANR.15}, and rotating nuclei \cite{AO.13}.

\begin{figure*}[ht]
\centering
\includegraphics[angle=0,width=14.0cm]{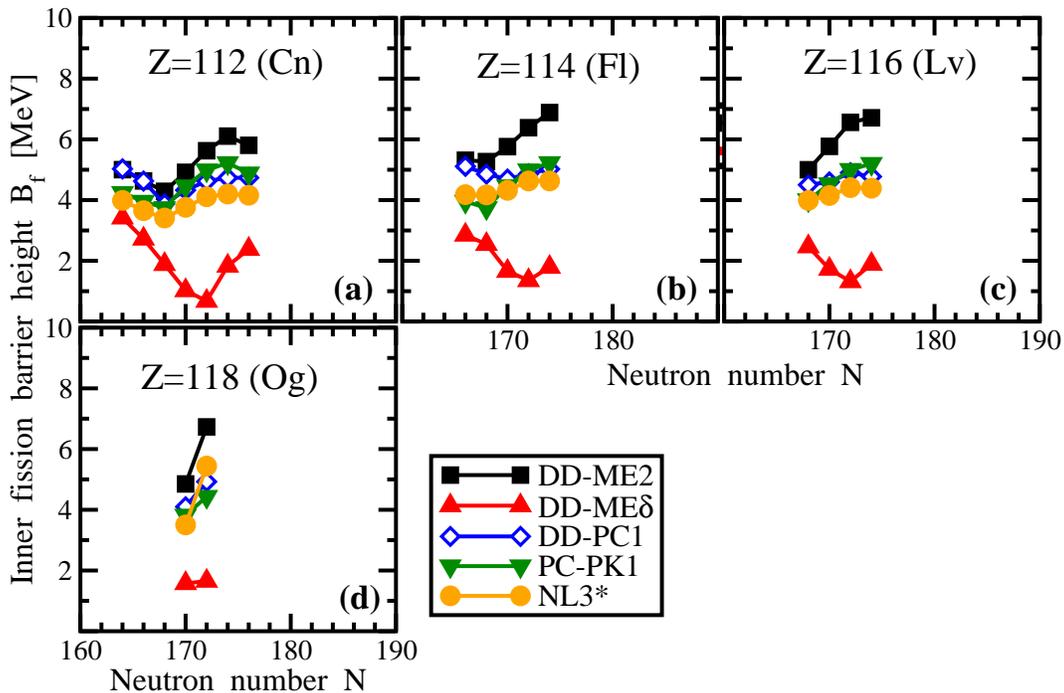}
\caption{(Color online) The heights of inner fission barriers in
selected nuclei as obtained in axially symmetric RHB calculations
with indicated CEDFs.}
\label{FB-SHE-axial}
\end{figure*}

 The paper is organized as follows. Section \ref{theory_details}
describes the details of the calculations. The results of global
investigation of inner fission barriers and related systematic
theoretical uncertainties within the axial RHB framework are
discussed in Sec.\ \ref{fission-axial}. Statistical uncertainties
in the description of fission barriers and potential energy curves
are investigated in Sec.\ \ref{stat-fission-axial}. Sec.\ \ref{fission-triax} is
devoted to the study of systematic uncertainties in the description
of the energies of fission saddles within the triaxial RHB framework.
In Sec.\ \ref{fission-barrier-dif-mod} we present a comparison of fission
barriers obtained in different models. Finally, Sec.\ \ref{concl} summarizes
the results of our work.

\section{Numerical details}
\label{theory_details}

  In the present manuscript, axially symmetric and triaxial
RHB frameworks are used for the studies of fission barriers
and the related theoretical uncertainties.

\begin{figure*}[ht]
\centering
\includegraphics[angle=-90,width=8.8cm]{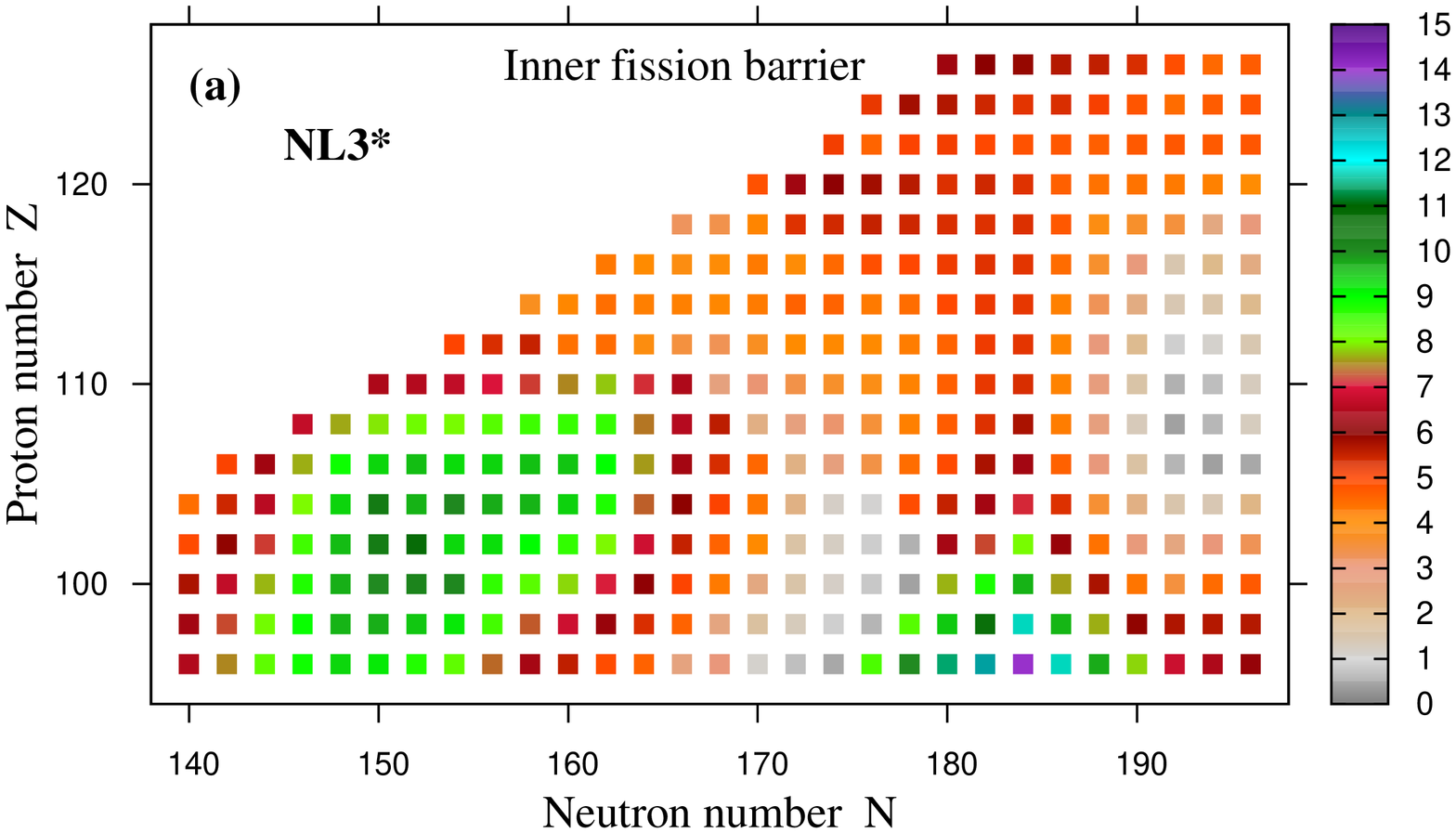}
\includegraphics[angle=-90,width=8.8cm]{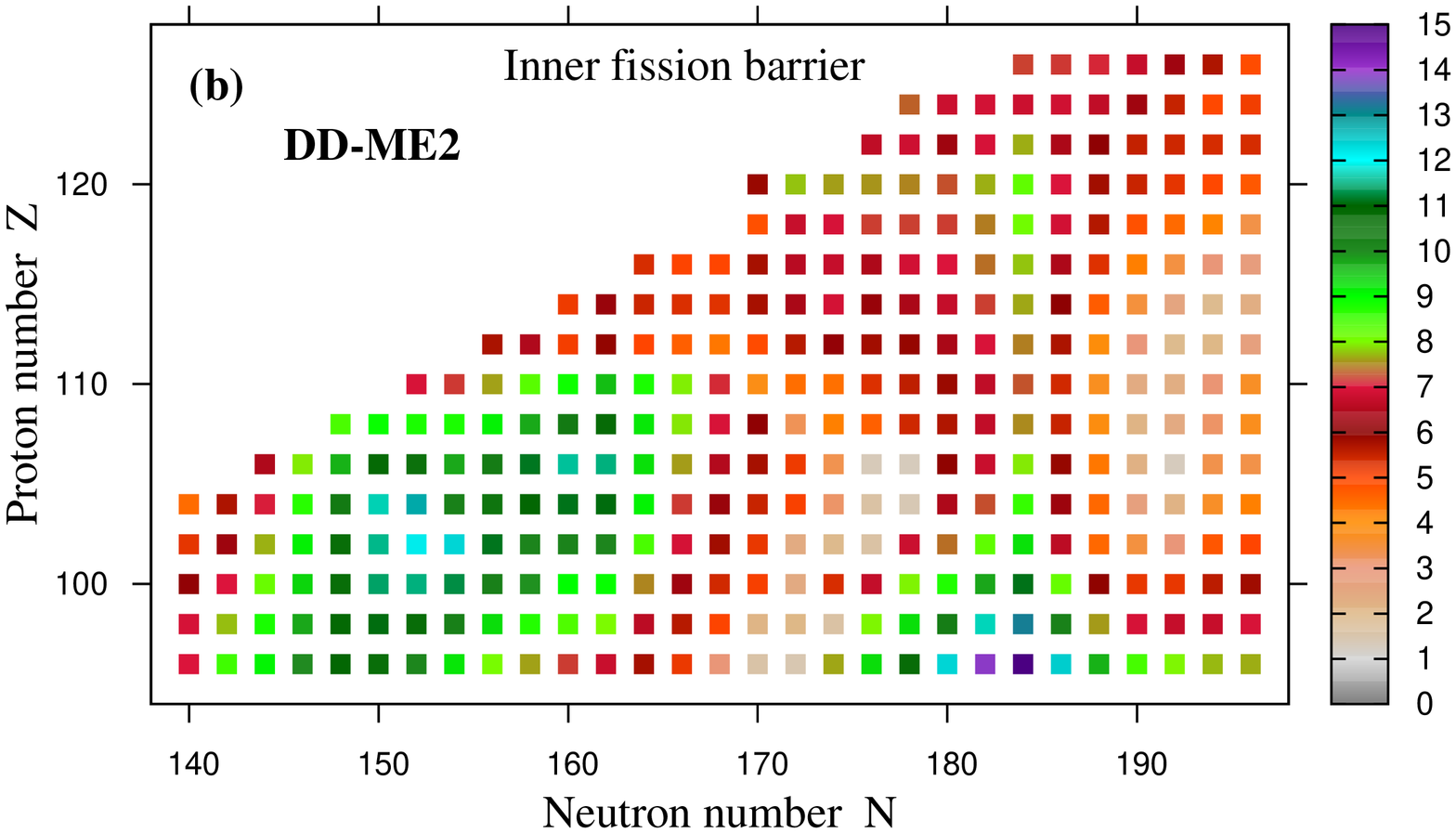}
\includegraphics[angle=-90,width=8.8cm]{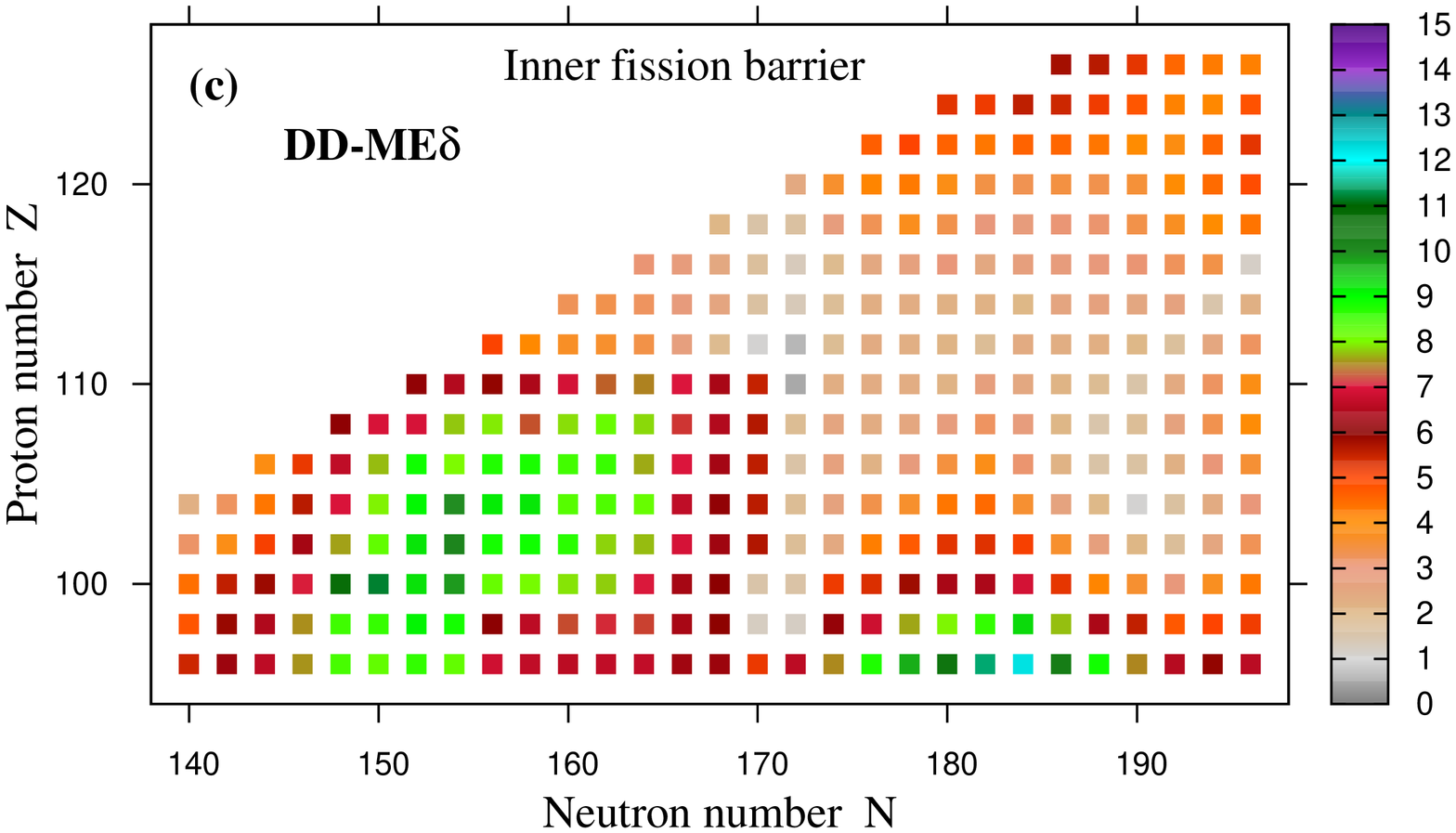}
\includegraphics[angle=-90,width=8.8cm]{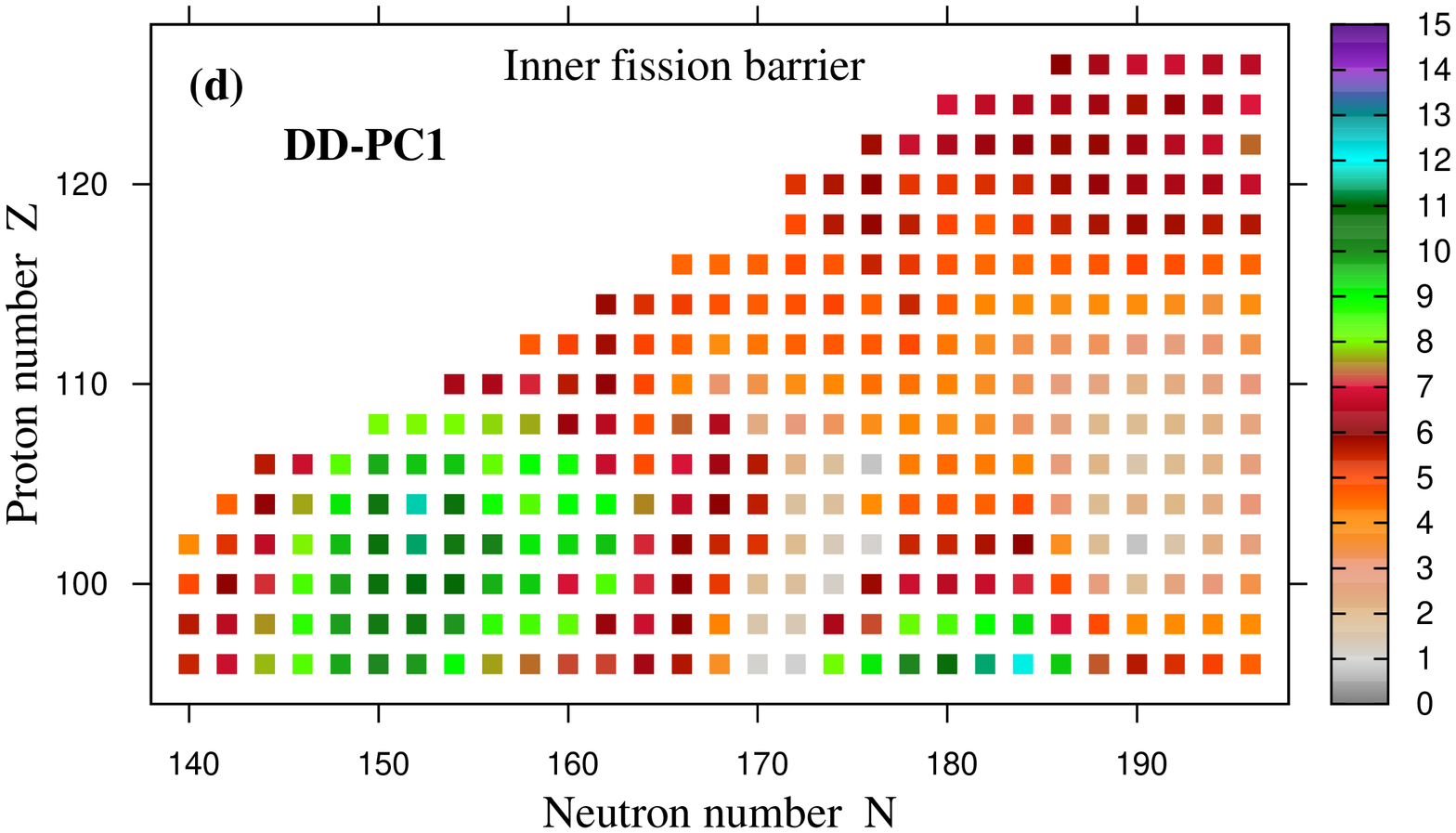}
\includegraphics[angle=-90,width=8.8cm]{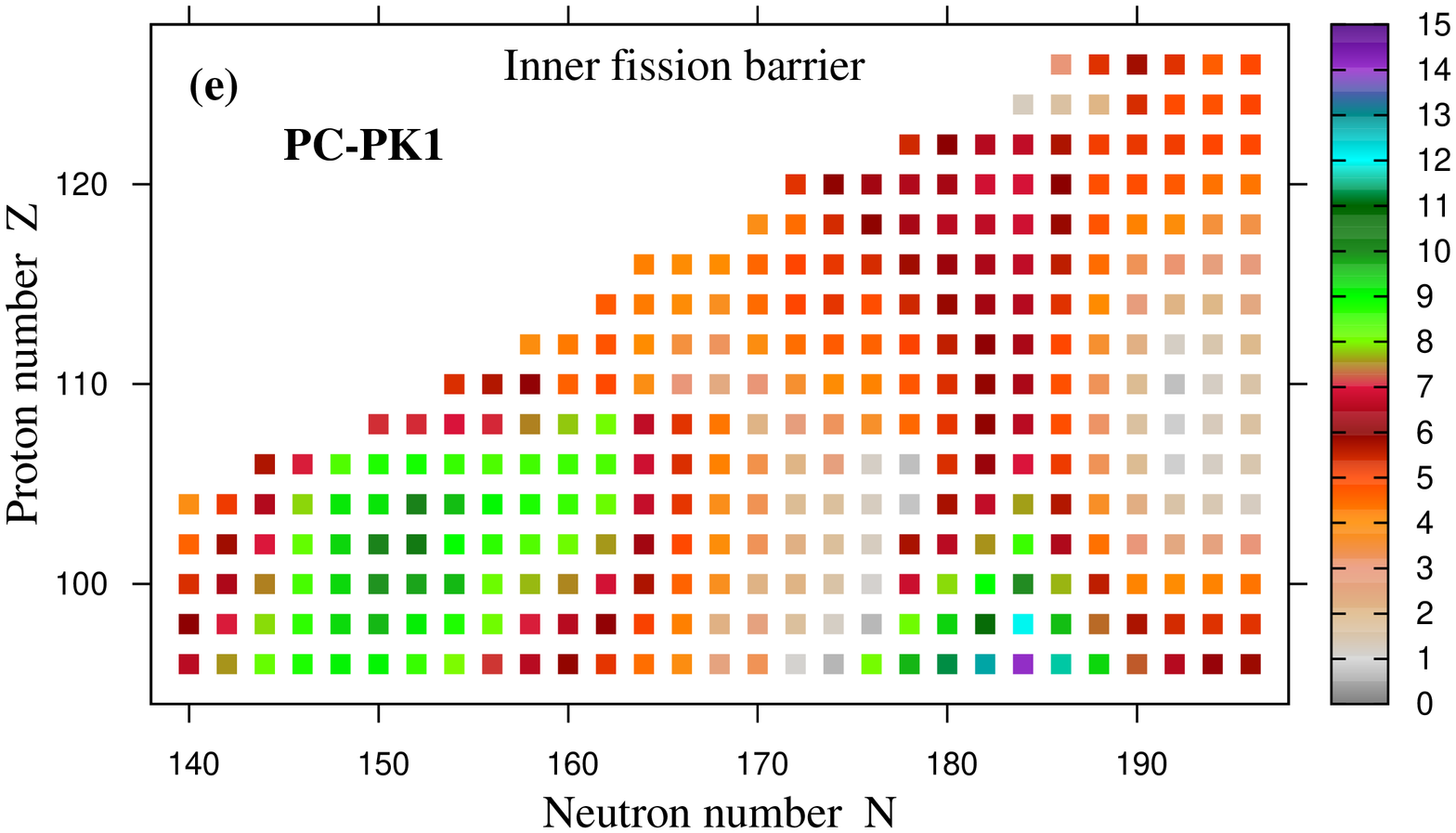}
\caption{(Color online) The heights of inner fission barriers (in MeV)
obtained in axially symmetric RHB calculations as a function of proton
and neutron number. The results of the calculations with the indicated CEDFs
are shown from the two-proton drip line up to $N=196$.}
\label{Sys_fiss_barr}
\end{figure*}

   First, the axially symmetric RHB framework is used for
systematic studies of all $Z=96-126$ even-even actinides and
SHEs from the proton-drip line up to neutron number $N=196$.
The proton-drip lines  for the different functionals are defined in
Refs.\ \cite{AARR.13,AARR.14,AANR.15}.
The details  of this formalism have been discussed in Secs.\ II-IV of Ref.\
\cite{AARR.14} and Sec.\ II of Ref.\ \cite{AARR.15}. Thus, we
only provide a general outline of the features specific for the
current RHB calculations. In these calculations, we
solve the RHB-equations in an axially deformed oscillator basis
\cite{PRB.87,GRT.90,RGL.97,Niksic2014_CPC185-1808,AAR.16}.
The truncation of the
basis is  performed in such a way that all states belonging to the
shells up to $N_F = 20$ fermionic shells and $N_B = 20$ bosonic shells are
taken into account.  As tested in a number of calculations with
$N_F=26$ and $N_B=26$, this truncation scheme provides sufficient
numerical accuracy.  For each nucleus the potential energy curve is
obtained in a large deformation range from $\beta_2=-1.0$ up to
$\beta_2=1.05$ in steps of $\beta_2=0.02$ by means of a constraint on
the quadrupole moment $Q_{20}$. Then, the correct ground state
configuration and its energy are defined; this procedure is especially
important for the cases of shape coexistence (see the discussion
in Ref.\ \cite{AANR.15}). The effect of the octupole deformation
on the binding energies of the ground states (and thus on the heights
of inner fission barriers) is also taken into account according to
the results obtained in Refs.\ \cite{AAR.16,AA.17}. Note that octupole
deformation in the ground states affects fission barriers and their
spreads only for the $Z\sim 92, N\sim 132$ and $Z\sim 96, N\sim 196$
nuclei.

 In order to avoid uncertainties connected with
the size of the pairing window, we use the separable form of the
finite range Gogny pairing interaction introduced by Tian et al \cite{TMR.09}.
As follows from the RHB studies with the CEDF NL3* of odd-even mass
staggerings, moments of inertia and pairing gaps, the Gogny D1S
pairing and its separable form
work well in the
actinides (Refs.\ \cite{AO.13,AARR.14,DABRS.15}). A weak dependence
of its pairing  strength on the CEDF has been observed in the studies of
pairing and rotational properties of deformed actinides in Refs.\ \cite{A250,AO.13},
of pairing gaps in spherical nuclei in Ref.\ \cite{AARR.14} and
of pairing energies in Ref.\ \cite{AARR.15}. Thus, in the present work, the same pairing
strength
is used also in the calculations with DD-PC1, DD-ME2,
DD-ME$\delta$, and PC-PK1. Considering the global character of this
study as well as the existing uncertainties in the extrapolation of
pairing from actinides (where experimental data could
be confronted with the results of calculations) to superheavy
nuclei, this is a reasonable choice.

\begin{figure*}[ht]
\centering
\includegraphics[angle=-90,width=11cm]{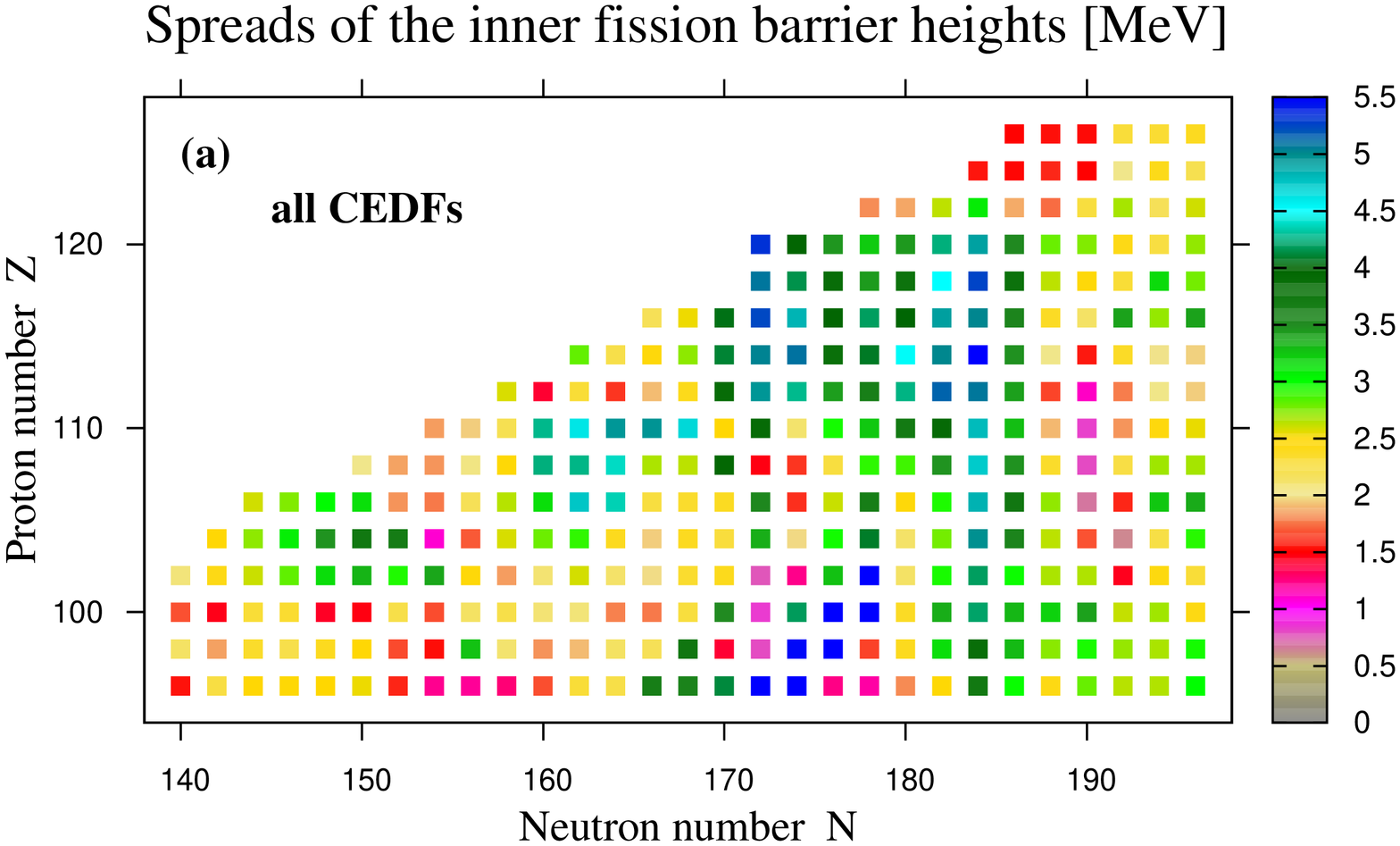}
\includegraphics[angle=-90,width=11cm]{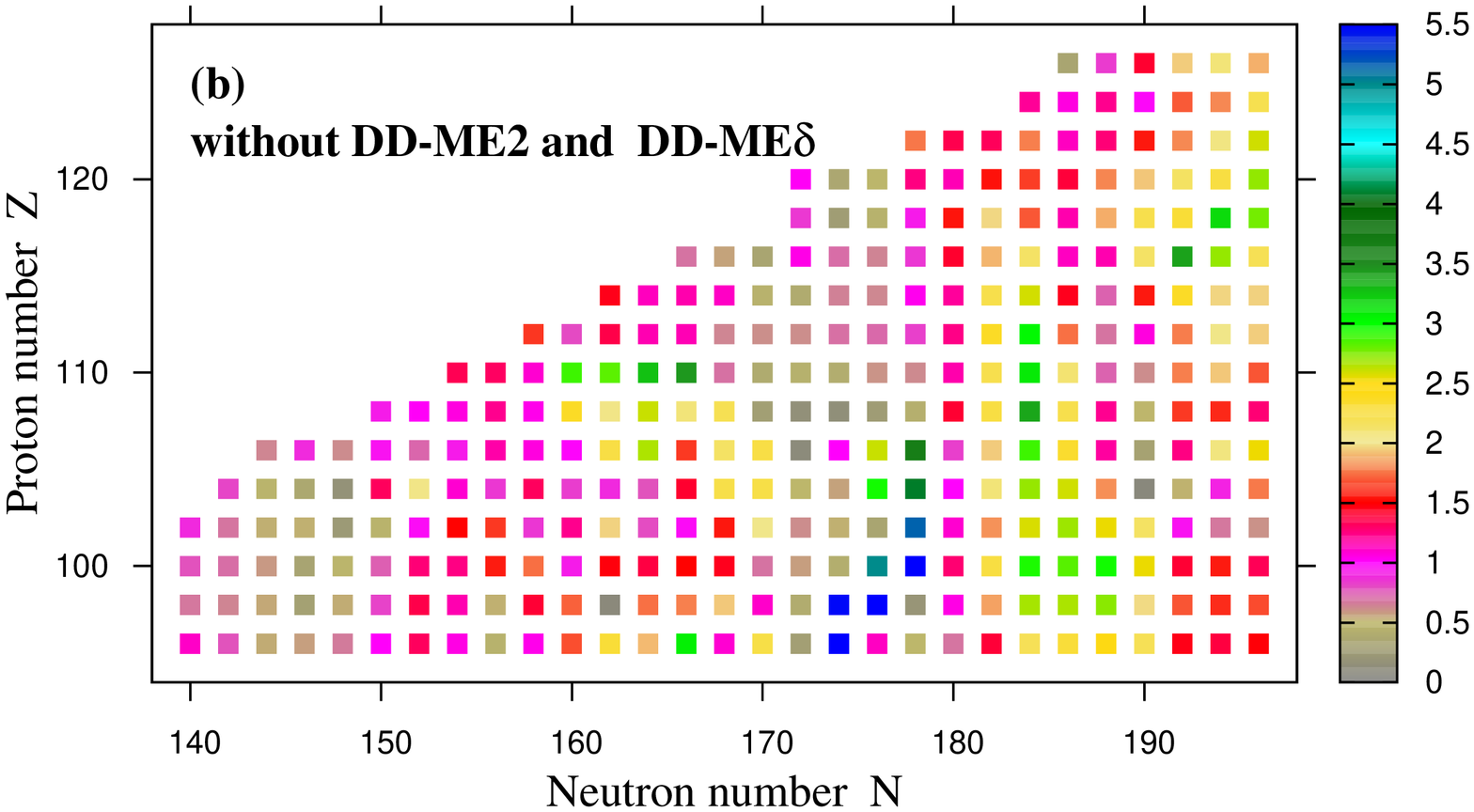}
\caption{(Color online) The spreads $\Delta E^B$ of the heights of inner fission
barriers as a function of proton and neutron number.
$\Delta E^B(Z,N) = |E^{B}_{max}(Z,N)-E^{B}_{min}(Z,N)|$, where, for
given $Z$ and $N$ values, $E^{B}_{max}(Z,N)$ and $E^{B}_{min}(Z,N)$ are
the largest and smallest heights of inner fission barriers obtained
with the set of functionals NL3*, DD-ME2, DD-ME$\delta$, DD-PC1, and
PC-PK1. Panel (a) shows the results for all five functionals, while
DD-ME2 and DD-ME$\delta$ are excluded in the results shown on panel
(b).}
\label{fission_spread}
\end{figure*}

\begin{figure*}[ht]
\centering
\includegraphics[angle=-90,width=8.8cm]{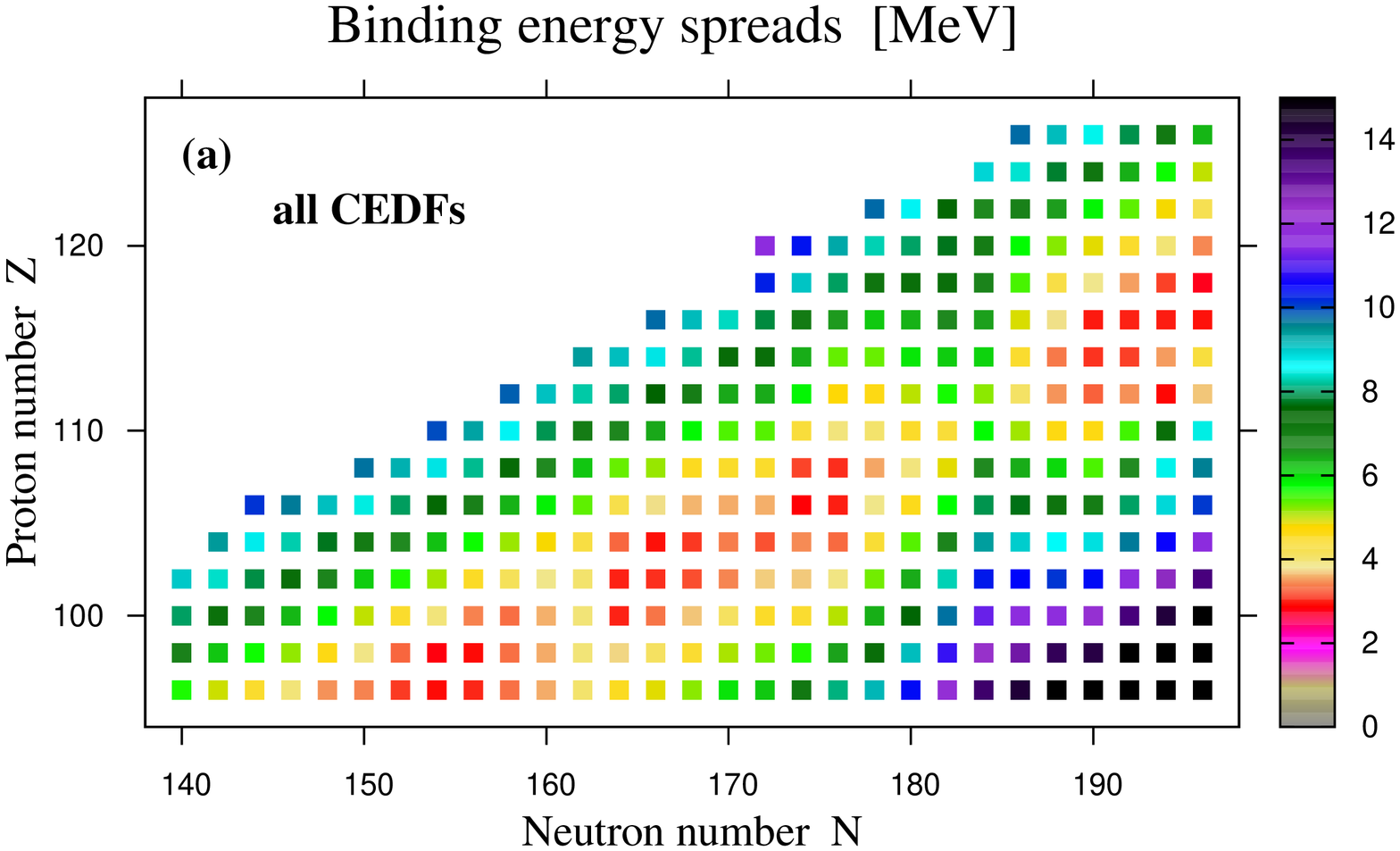}
\includegraphics[angle=-90,width=8.8cm]{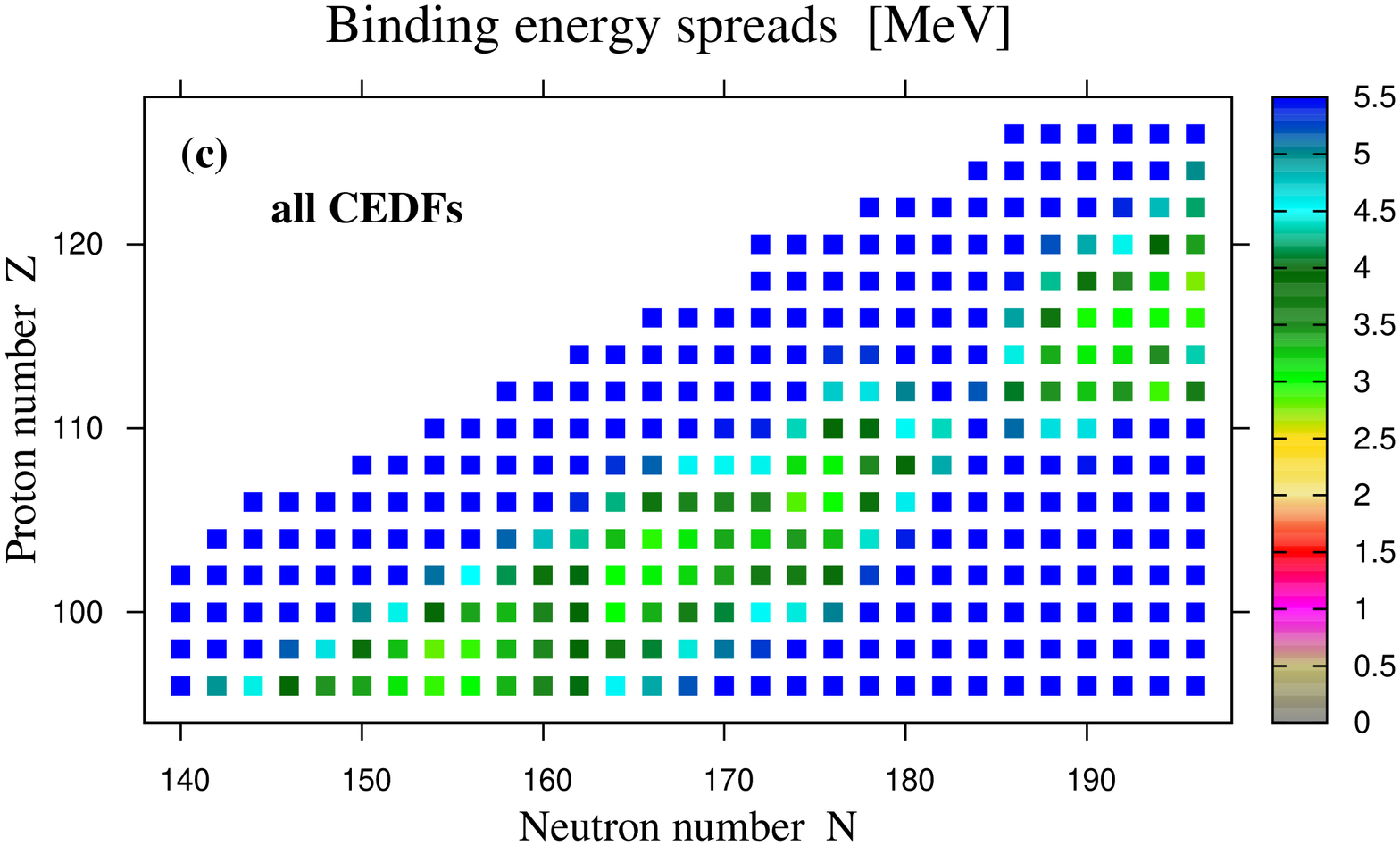}
\includegraphics[angle=-90,width=8.8cm]{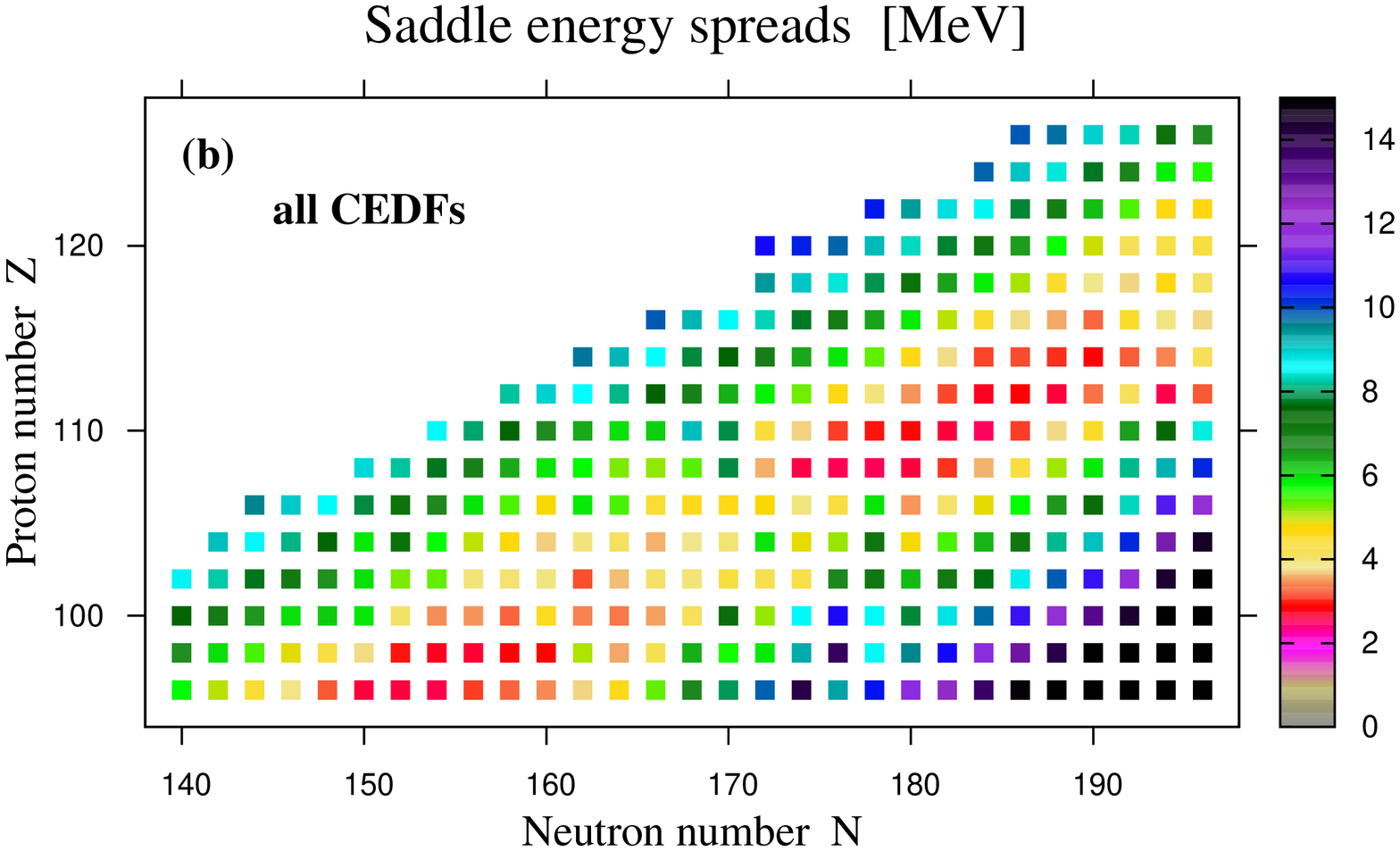}
\includegraphics[angle=-90,width=8.8cm]{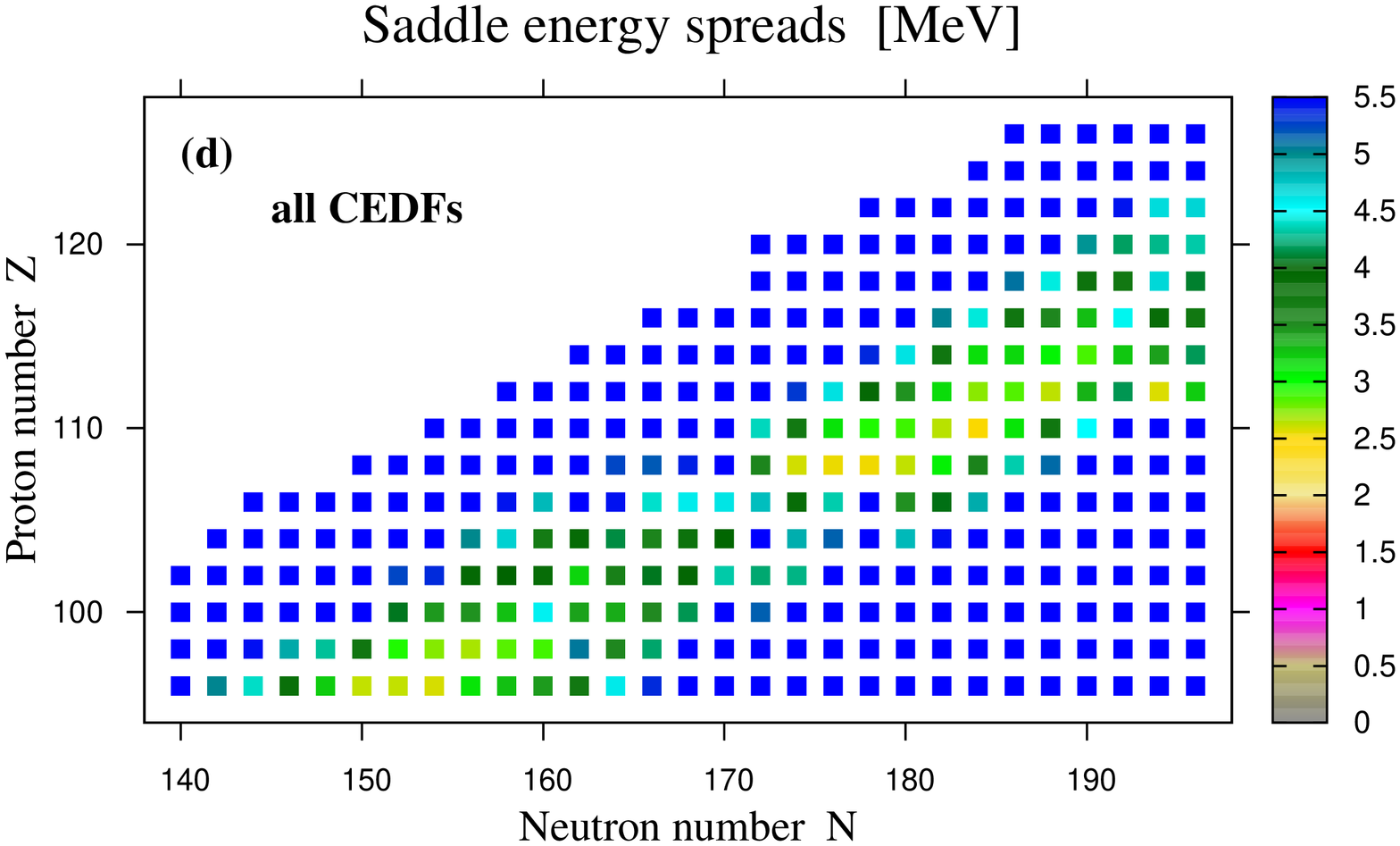}
\caption{(Color online) The spreads in the binding energies of the
ground states (panel (a)) and the energies of saddle points 
(panel (b)) as obtained with five employed CEDFs. Panels (c) and (d) 
show the same results but with the energy colormap used in Fig.\ 
\protect\ref{fission_spread}.}
\label{BE-saddle-spread-all}
\end{figure*}

\begin{figure*}[ht]
\centering
\includegraphics[angle=-90,width=8.8cm]{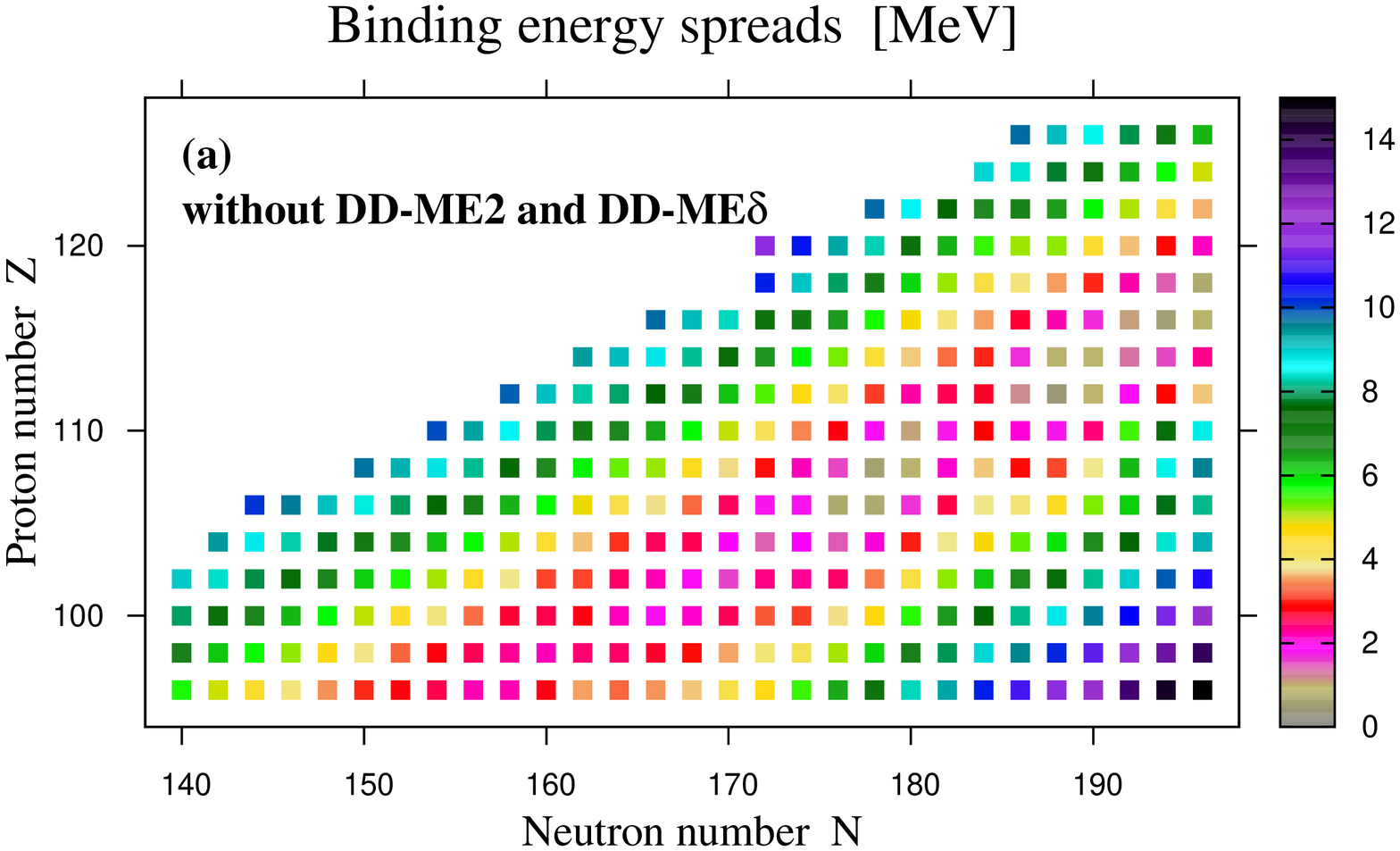}
\includegraphics[angle=-90,width=8.8cm]{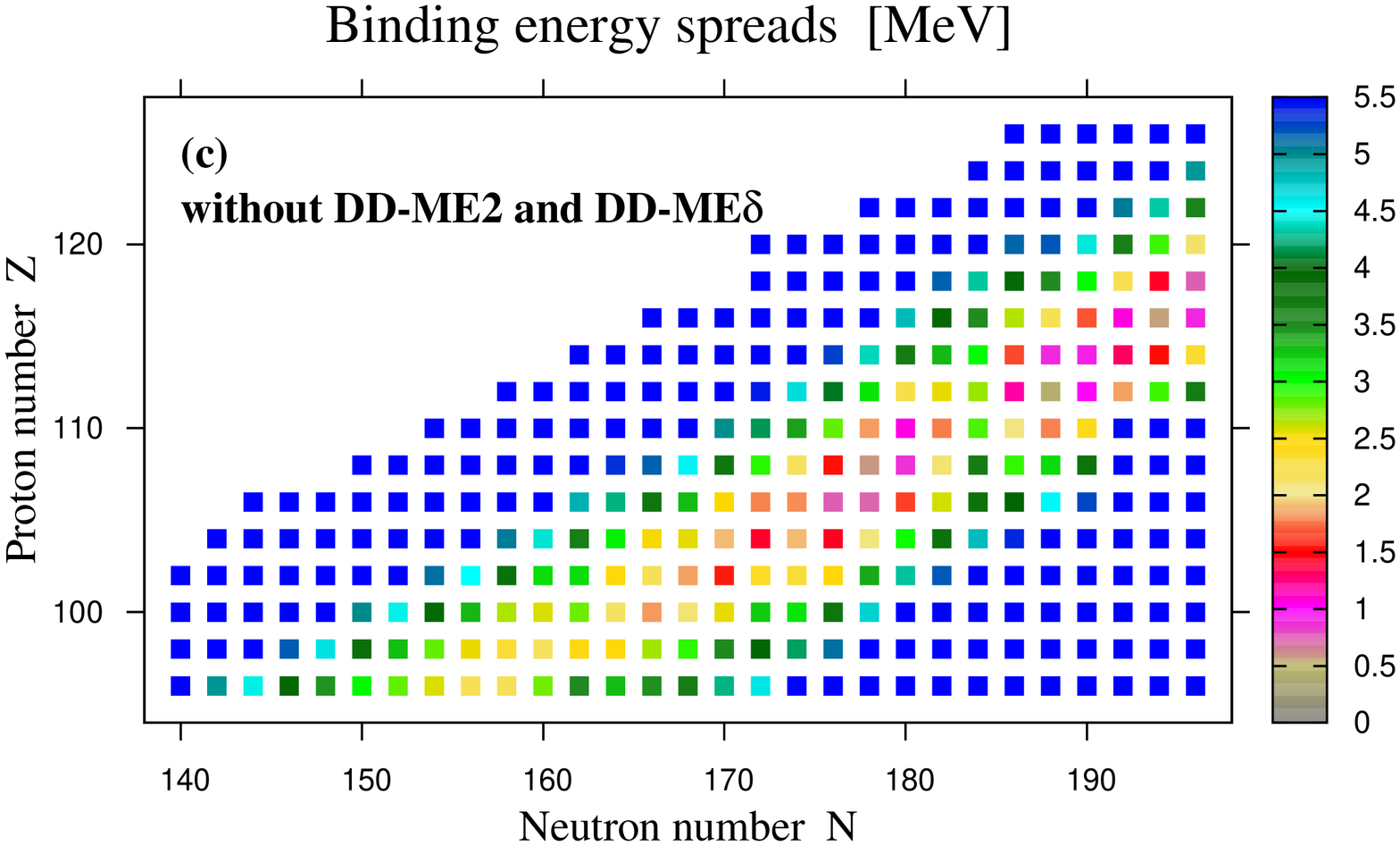}
\includegraphics[angle=-90,width=8.8cm]{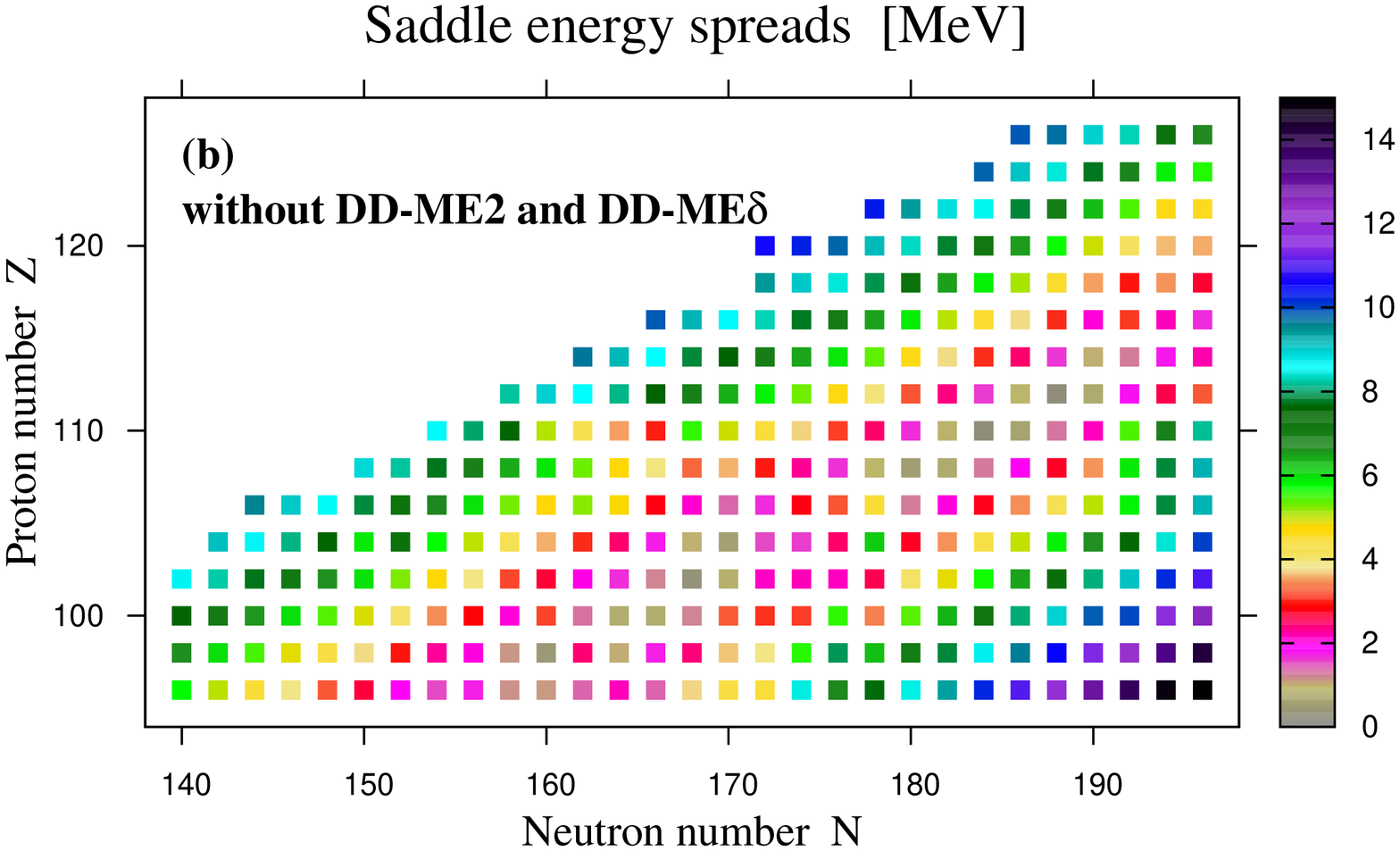}
\includegraphics[angle=-90,width=8.8cm]{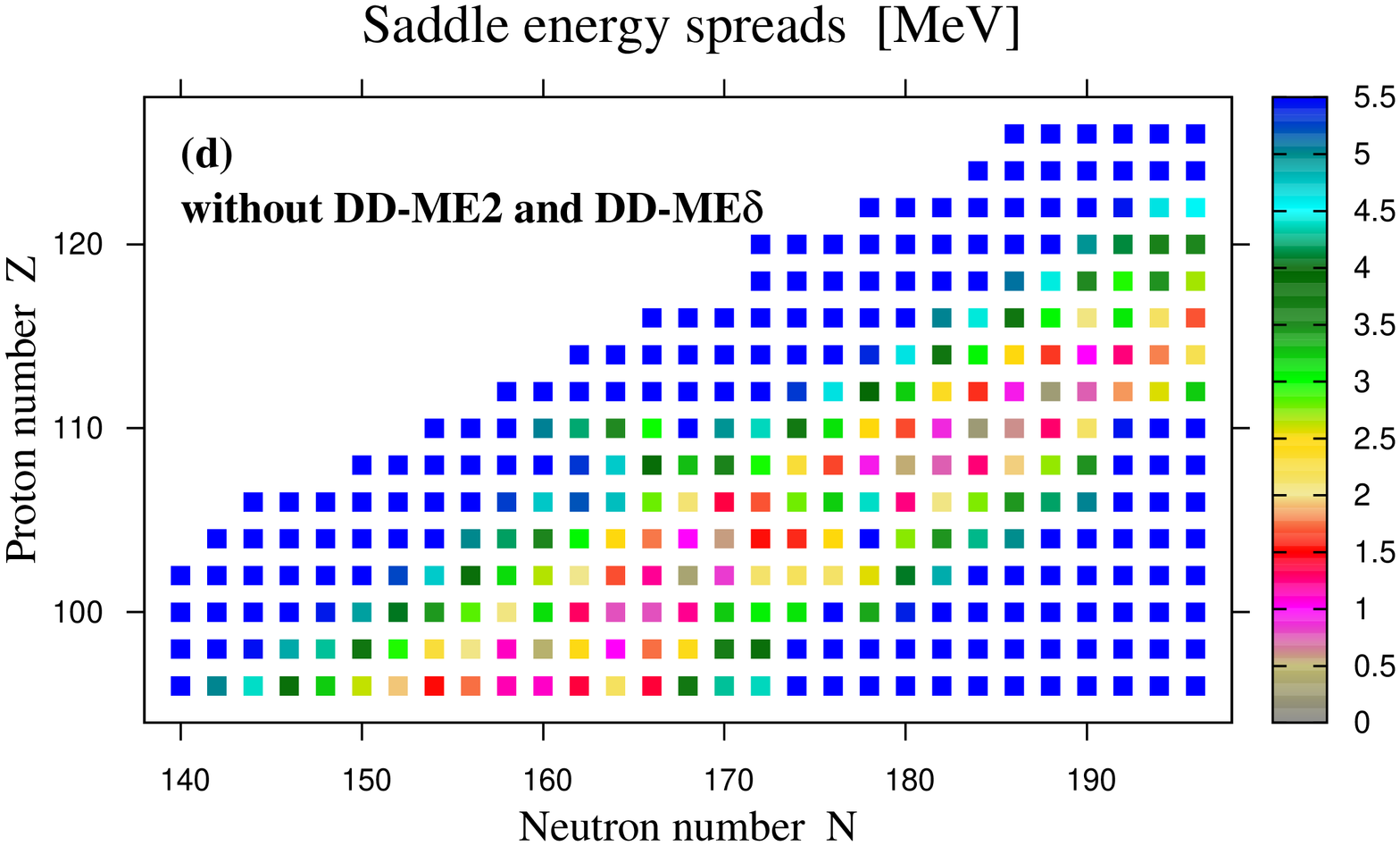}
\caption{(Color online) The same as Fig.\ \protect\ref{BE-saddle-spread-all} but
for  the case when DD-ME2 and DD-ME$\delta$ CEDFs are excluded from consideration.}
\label{BE-saddle-spread-lim}
\end{figure*}

  As a next step, we perform triaxial RHB (TRHB) calculations
in a parity conserving cartesian oscillator basis
\cite{CRHB,Niksic2014_CPC185-1808}
using the same pairing and the same set of the functionals.
However, such calculations are enormously time-consuming.
Therefore, they cannot be carried out on the same global scale as axial
RHB calculations. As a result, we restricted the TRHB studies to a
selected set of the $Z=112-120$ nuclei. These nuclei are located
mostly in the region where extensive
experimental studies have either  been already performed or will be
performed in a foreseeable future. Even then the calculations of
full potential energy surfaces (PES) are numerically prohibitive
for the $N_F=20$ fermionic basis. However, the topology of the PESs obtained
in the TRHB calculations with the truncation of the fermionic basis at $N_F=16$
and $N_F=20$ is the same. Thus, full PESs have been calculated only with the
$N_F=16$ fermionic basis. These results define the positions in the
deformation plane and the energies of axial and triaxial saddles.
Afterwards, they are corrected for the $N_F=20$ fermionic basis by performing
the TRHB calculations with the $N_F=20$ fermionic basis in the spherical/normal
deformed minimum and at few grid points near the saddles.

\section{Global investigation of inner fission barriers and
related systematic theoretical uncertainties in the axial RHB
calculations.}
\label{fission-axial}

  Fig.\ \ref{FB-SHE-axial} compares the heights of inner fission
barriers obtained in axially symmetric RHB calculations with the
five functionals. We show only results for nuclei in which the
lowest saddle is axially symmetric in the systematic triaxial
RMF+BCS calculations with NL3* of Ref.\ \cite{AAR.12}. One can
see that NL3*, DD-PC1 and PC-PK1, which successfully describe
experimental fission barriers in the actinides
\cite{AAR.10,AAR.12,LZZ.12,PNLV.12,LZZ.14}, give similar results
for the heights of inner fission barriers. On the other hand, the
fission barriers produced by DD-ME2 are always at the upper end.
This may be a generic feature of this functional since it produces
also in $^{236}$U  and $^{240}$Pu inner fission barriers which
are higher than those of NL3* and DD-PC1 \cite{AAR.12}. The functional
DD-ME$\delta$ produces unrealistically low fission barriers (see
Sec.\ III in Ref.\ \cite{AAR.12} for a discussion of the inner
fission barriers in SHEs).

  The global behavior of the inner fission barrier heights in the
region of superheavy nuclei is shown in Fig.\ \ref{Sys_fiss_barr}
for all five employed functionals. Again the highest fission barriers
are provided by DD-ME2 and the lowest by DD-ME$\delta$.

  The employed functionals can be split into two groups \cite{AANR.15}.
The first group, consisting of NL3*, DD-ME2 and PC-PK1, predicts bands
of spherical SHEs in the $(Z,N)$ plane centered around the $Z=120$ and $N=184$
lines. The second group includes DD-ME$\delta$ and DD-PC1 and it
does not predict spherical SHE in the vicinity of above mentioned
particle numbers. The
impact of the proton and neutron spherical shell gaps at $Z=120$ and
$N=184$ is clearly visible for NL3*, DD-ME2, and PC-PK1; there is
a substantial increase of the inner fission barrier heights around
these numbers. In contrast, no such effect is seen in the
calculations with DD-ME$\delta$ and DD-PC1. For NL3* and PC-PK1,
the heights of the inner fission barriers are lowered around
$Z\sim 100, N\sim 172$ and $Z\sim 108, N\sim 194$. Similar
regions of reduced inner fission barrier heights could be found
also for the other functionals but they are centered around different
combinations of proton and neutron numbers.

  The spreads in the predictions of inner fission barrier heights
are shown for all five employed functionals in Fig.\ \ref{fission_spread}a.
One can see that in the actinides ($Z\leq 100, N\leq 164$) these spreads
are typically smaller than 2.5 MeV. Note that in this mass region theoretical
uncertainties in the prediction of the ground state deformations are very
small  (see Refs.\ \cite{AARR.14,AANR.15}). However, the $\Delta E^B$
spreads drastically increase in the $Z=112-120, N=170-186$ region where they
range from 3.5 MeV up to 5.5 MeV. To a large extent this region coincides
with the region where the uncertainties in the predictions of the ground state
deformations are substantial (see Fig. 8 in Ref.\ \cite{AANR.15}). This
clearly suggests that in this region the uncertainties in the fission barrier
heights are strongly affected by the uncertainties in the ground state
deformations. A similar enhancement of the $\Delta E^B$ spreads is seen in the nuclei
around $Z\sim 98, N\sim 174$. However, the differences in the predictions of the
ground state deformations play here a minor role since they are almost the
same for all functionals (see Fig. 8 in Ref.\ \cite{AANR.15}). Theoretical
$\Delta E^B$ spreads decrease for $N\geq 186$; here they are typically less
than 3 MeV with only a few nuclei characterized by higher spreads of around 4 MeV.

The above discussed impact of the uncertainties in the calculated
deformations on the spreads of inner fission barrier heights can be
understood in the following way. The inner fission barrier height
is the difference between the energies of the saddle and ground state.
However, these two points in the potential energy curve have
different deformations and thus substantial differences in the
underlying shell structure. This leads to different spreads
of the binding energies in the ground states and saddles which are
compared in Fig.\ \ref{BE-saddle-spread-all}. Minimum spreads in
these energies appear in the band of the nuclei which is shown
in yellow and red  colors (Figs.\ \ref{BE-saddle-spread-all}a and b).
These spreads increase on going away from this band of the nuclei;
this is caused by different isovector properties of employed functionals
(see discussion in Ref.\ \cite{AA.16} for more details). Let focus our
discussion on this yellow/red band of the nuclei.  Due to different underlying
shell structure at the ground state and saddle point, the minima of the spreads
(shown by red/reddish colors) in the binding energies are localized in
different $(Z,N)$ regions at the ground state and saddle point.
Indeed, at the ground state the increase of the spreads in binding
energies takes place near neutron numbers which correspond to the
shell gaps in the single-particle spectra, namely, near deformed
$N=162$ shell closure and especially near $N\sim 184$ which
corresponds to spherical shell closure in some functionals
(Fig.\ \ref{BE-saddle-spread-all}a). The situation is different at
the saddle point where the increase of the spreads in binding
energies appears in wide regions near $N\sim 168$ and $\sim 196$
(Fig.\ \ref{BE-saddle-spread-all}b). These effects become even
more visible when the colormap of Fig.\ \ref{fission_spread}
is used in Figs.\ \ref{BE-saddle-spread-all}c and d; this is
done for simplicity of the comparison of these two figures. The
features of the fission barrier spreads which are visible in
Fig.\ \ref{fission_spread} (and especially their increase near
$N\sim 184$) are consequences of the ones seen in Fig.\
\ref{BE-saddle-spread-all}.

  The benchmarking of the functionals to experimentally known fission
barriers in the actinides allows to reduce theoretical spreads in their
heights for unknown nuclei. This is illustrated in Fig.\ \ref{fission_spread}b
where only the NL3*, DD-PC1 and PC-PK1 functionals are used in the definition
of the theoretical spreads. Again the source of this reduction could
be traced back to the reduction of the fluctuations in binding energy
spreads for the ground states and saddles in the direction along the
direction of minimum spreads [compare Figs.\ \ref{BE-saddle-spread-lim}
and \ref{BE-saddle-spread-all}].
These functionals successfully describe experimental
fission barriers in the actinides \cite{AAR.10,AAR.12,LZZ.12,PNLV.12,LZZ.14}.
One can see that the use of only these functionals reduces theoretical
uncertainties in the inner fission barrier heights for the $N\leq 180$ nuclei
typically to less than 2 MeV; only in few nuclei around $Z=110,
N\sim 164$ and $Z\sim 110, N\sim 176$ these uncertainties are
higher reaching 4 and 5.5 MeV respectively. However, these
uncertainties increase by roughly 1 MeV for the nuclei with
$N\geq 182$. It is also important to mention that theoretical
spreads in the inner fission barrier heights do not form a smooth
function of proton and neutron numbers; there is always a random
component in their behavior.

\section{Statisitical uncertainties in the description of fission
barriers and potential energy curves}
\label{stat-fission-axial}

  As discussed in the introduction, there are statistical uncertainties
in the description of physical observables in addition to the systematic
ones which for the saddles of inner fission barriers are quantified in the
previous section and in Secs.\ \ref{fission-triax} and \ref{fission-barrier-dif-mod}
below. The description of the statistical uncertainties for fission barriers
and potential energy curves follows the formalism presented in Ref.\
\cite{DNR.14}. Its major details are shortly outlined below.

  For a model having $N_{par}$ adjustable parameters
${\bf p}=(p_1, p_2, ..., p_{N_{par}})$ the normalized objective function
is defined as
\begin{eqnarray}
\chi^2_{norm}({\bf p})=\frac{1}{s}\sum_{i=1}^{N_{type}} \sum_{j=1}^{n_i} \left( \frac{O_{i,j}({\bf p})-O_{i,j}^{exp}}
{\Delta O_{i,j}} \right)^2
\label{Ksi}
\end{eqnarray}
where
\begin{eqnarray}
s=\frac{\chi^2({\bf p}_0)}{N_{data}-N_{par}}
\end{eqnarray}
is a global scale factor (Birge factor \cite{Birge.32})  defined at the
minimum of the penalty function (optimum parametrization ${\bf p}_0$)
which leads to the average $\chi^2 ({\bf p}_0)$ per degree of freedom
equal to one \cite{DNR.14} and
\begin{eqnarray}
N_{data}= \sum_{i=1}^{N_{type}}n_i
\end{eqnarray}
is the total number of data points of different types.
Here, $N_{type}$ stands  for the number of different data types.
The calculated and experimental/empirical values of the physical
observable $j$ of the $i-$th type are represented by $O_{i,j}({\bf p})$
and $O^{exp}_{i,j}$, respectively. $\Delta O_{i,j}$ is the adopted error
for the physical observable $O_{i,j}$.

The acceptable functionals are defined by the condition \cite{Stat-an,DNR.14}
\begin{eqnarray}
\chi^2_{norm} ({\bf p})  < \chi^2_{norm}({\bf p}_0) + 1 .
\label{cond}
\end{eqnarray}
This condition specifies the 'physically reasonable' domain around the
minimum ${\bf p}_0$ in which the parametrization ${\bf p}$ provides a
reasonable fit and thus can be considered as acceptable. For a given
original functional the set of the $M$ functional variations
$\left[ {\bf p}_1, {\bf p}_2, ..., {\bf p}_M \right]$  has been defined
in Ref.\ \cite{AA.16-prep}; note that this set also includes the original
functional.

  For this set of the functional variations the potential energy curves
of the $^{296}$112 nucleus have been calculated in the axial RHB framework
in the deformation range $\beta_2=0.0-1.05$ with a step of
$\Delta \beta_2=0.02$. Then, the energies of these potential energy curves
were redefined  with respect of the energy of their spherical or near
spherical minimum. As a result, the energy of the minumum becomes equal
to zero. Finally, the mean values of the energy
\begin{eqnarray}
\bar{E}(\beta_{2,i})=\frac{1}{M} \sum_{k=1}^{M} E_k(\beta_{2,i})
\label{e-mean}
\end{eqnarray}
and their standard deviations
\begin{eqnarray}
\sigma_{E} = \sqrt{
\frac {1}{M} \sum_{k=1}^{M} [ E_k(\beta_{2,i}) -\bar{E} (\beta_{2,i}) ]^2
}
\label{e-stand}
\end{eqnarray}
have been calculated for this set of potential energy curves at the $i$-th value of
the deformation. Note that the standard deviations serve here as a measure of the
statistical uncertainty.

The set of the $M$ functional variations defined by the condition
of Eq.\ (\ref{cond}) is represented by two thousand ``reasonable'' functionals
\cite{AA.16-prep}. Note that they are defined with respect of the ground
state experimental data on 12 spherical nuclei and some empirical data on
nuclear matter properties. However, because of numerical restrictions we use
the subset of 500 randomly selected functionals in the calculations of potential
energy curves and quantities defined in Eqs.\ (\ref{e-mean}) and (\ref{e-stand}).
This number of the functionals is sufficient for a reliable definition of the mean
values of the energy and their standard deviations. Indeed, the comparison of the
results obtained with 250 and 500 functionals reveals very little difference which
strongly suggests that fine details related to statistical properties of the
quantities of interest are already imprinted in relatively small number (few
hundred) of the functional variations.

  The selection of the $^{296}$112 nucleus has been guided by the requirement
to avoid large shape changes in the ground state with the variation of the
functional. Indeed, this nucleus is spherical in the ground state with a well
pronounced minimum in the parabola-like potential energy surface. As a consequence,
with exception of a few functional variations, the ground state is spherical
in the RHB calculations. On the contrary, larger shape changes in the ground
state with the variation of the functional are expected  in many nuclei of the
region of interest since they are transitional in the ground state with soft
potential energy surfaces (see Refs.\ \cite{AAR.12,AANR.15}). In such nuclei,
statistical theoretical uncertainties in the evolution of the energy with
deformation in potential energy curves are expected to be polluted by the
variations in the ground state properties.

\begin{figure}[ht]
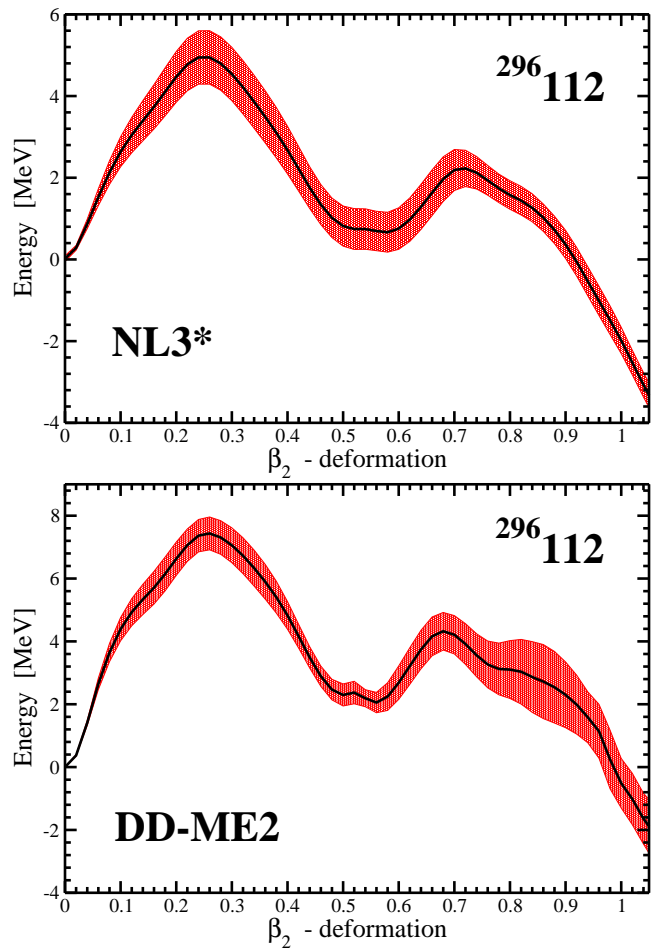

\centering
\includegraphics[angle=0,width=8.5cm]{fig-6-a.eps}
\includegraphics[angle=0,width=8.5cm]{fig-6-b.eps}
\caption{(Color online) Statistical uncertainties in the deformation
energy curves of the $^{296}$112 nucleus. The mean potential energy
curve is shown by a solid line. The red colored region shows the
standard deviations in energy.}
\label{FB-stat-uncert}
\end{figure}

 Statistical uncertainties in the deformation energy curves are shown
in Fig.\ \ref{FB-stat-uncert}. One can see that in the case of the
NL3* functional they are small in
the vicinity of the spherical minimum but then increase with increasing
deformation. They become especially pronounced in the vicinity of
the inner and outer saddles and in the region of the superdeformed (SD) minimum.
The $\sigma_{E}$ values have a maximum at the inner saddle where they are
close to 0.7 MeV. They are smaller at the superdeformed minimum and the outer
saddle where they are close to 0.5 MeV. It is interesting that
statistical uncertainties decrease substantially above the outer
fission barrier; here $\sigma_{E}$ values are around 0.35 MeV.

  Their behavior is quite different for DD-ME2. For this functional the
statistical uncertainties in the deformation range $\beta_2=0.0-0.6$
are by approximately factor 1.5 smaller than those obtained with NL3*.
However, they increase with increasing deformation and reach a maximum at
$\beta_2 \sim 0.85$. With further increase of deformation, they decrease
and finally stabilize above $\beta_2\sim 0.95$.

 It is likely that the increased theoretical uncertainties in the
region of quadrupole deformations $\beta_2 =0.1-0.8$ in the case
of NL3* and around $\beta_2 \sim 0.85$ in the case of DD-ME2 are due
to the underlying single-particle structure. The variations of the functional
lead to modifications of the single-particle energies as well as to
changes in the sizes of the superdeformed shell gaps and the
single-particle level densities at the saddles and the SD minimum.
This in turn leads to substantial variations in the shell
correction  energies.
The DD-ME2 functional provides a clear example. Indeed, the increase
of statistical uncertainties around $\beta_2\sim 0.85$ is due to the
fact that an additional hump develops in the potential energy curve at this
deformation in a number of  the functional variations. This fact
could only be explained by the underlying shell structure.
In addition, the reduced statistical uncertainties at larger
deformations and their stabilization strongly support the impact
of the underlying single-particle structure on the statistical uncertainties.
This is because the shell gaps
at hyper- and higher deformations are smaller than at
superdeformation (see Refs.\ \cite{SRN.72,AA.08} and references quoted therein).
As a consequence, the changes in the single-particle structure caused
by the functional variations have a smaller impact on the shell correction
energy.

 So far the statistical uncertainties in the deformation energy
curves have been investigated only in Skyrme DFT. They
have been studied in Ref.\ \cite{MSHSWN.15} for the nucleus $^{240}$Pu with the functional UNEDF1
functional  and in Ref.\ \cite{BKRRSW.15} for the nucleus $^{266}$Hs
with the functionals SV-min and SV-bas.
In $^{240}$Pu, the statistical
uncertainties  increase on going from the normal deformed minimum to
higher deformations and they become
especially large after the second fission barrier. The later observation is
in contradiction with our results which show a decrease and a stabilization
of statistical uncertainties after the second fission barrier. The reasons
for such a difference are not clear. They may be related to a different
choice of the nuclei. Differences in the analysis also contribute.
To avoid the use of the emulator, we restricted our consideration to axially
symmetric shapes. On the contrary, the authors of Ref.\ \cite{MSHSWN.15}
replace the exact DFT model by a Gaussian process response surface (GPPS)
which they `train'' to a restricted set of 200 DFT computations. Although
this simplified approach allows to take into account also triaxiality
and octupole deformation, it  is not clear how well GPPS reproduces the
energies for very complicated topologies of the potential energy surfaces
(see for example, Figs.\ \ref{PES-112-184} and \ref{PES-120-300} below) for
functional variations not included into the ``training'' set. The analysis
of $^{266}$Hs in Ref.\ \cite{BKRRSW.15} is restricted to the vicinity of the
inner fission barrier. Statistical uncertainties obtained with the SV-bas
(SV-min) functional are close to (larger than) those obtained in our
analysis of $^{296}$112.

\section{Systematic theoretical uncertainties in the description
of inner fission barriers for triaxial RHB calculations.}
\label{fission-triax}

   Not in all cases the axial saddle is the lowest in energy. The
systematic investigation of the heights of inner fission barriers in
superheavy nuclei performed within the RMF+BCS approach with the NL3*
CEDF in Ref.\ \cite{AAR.12} has revealed that triaxial deformation
lowers the heights of the inner fission barriers in a number of
nuclei; this is especially pronounced in the vicinity of particle numbers $Z=120$ and
$N=184$ (see Table V in Ref.\ \cite{AAR.12}). Thus,
the axial RHB calculations provide an upper limit for the inner
fission barrier heights.

   In general, triaxial deformation has to be included into the
calculations for a more realistic estimate of the heights of inner
fission barriers which can be used for the comparison with experiment.
However, such a study requires tremendous computational power.  This
is also illustrated by the fact that within the covariant and Gogny DFTs,
so far,
only a limited set of superheavy nuclei has been studied in the triaxial
Hartree-Bogoliubov framework \cite{PNLV.12,PNV.13,WE.12}. The
computational challenge becomes especially large in the case of
the analysis of systematic theoretical uncertainties because the same
nucleus has to be calculated within the TRHB framework for several CEDFs.
Thus, a full global analysis of theoretical uncertainties similar to the
one presented in Sec.\ \ref{fission-axial} in the axial RHB framework
is, at present, beyond the reach of available computational facilities. As a result,
we concentrate here on the selected set of the $Z=112-120$ superheavy nuclei which
will be in the focus of experimental studies within the next decades.
They are shown in Fig.\ \ref{fission_spread_triax} below. In the
selection of nuclei we focus on the nuclei in which the triaxial
saddle is expected to be the lowest in energy in the region of
interest. According to systematic studies in the RMF+BCS framework
with the CEDF NL3* of Ref.\ \cite{AAR.12}, these are the nuclei in
the vicinity  of the $Z=120$ and $N=184$ lines. On the contrary,
the axial saddles are the lowest in energy in the nuclei which are
away from these lines. For example, this takes place for $N\leq 174$
in the $Z=112,114,116$ nuclei (see Ref.\ \cite{AAR.12}). Triaxial RHB
calculations for the $(Z=112,N=164)$, $(Z=112, N=172)$, $(Z=114, N=166)$
and $(Z=114,N=172)$ nuclei (these nuclei are seen on the left side of Fig.\
\ref{fission_spread_triax}a) confirm this observation of Ref.\ \cite{AAR.12}
for all CEDFs employed in the present manuscript. We will try to establish
(i) how theoretical systematic uncertainties obtained in axial RHB
calculations will be modified when triaxiality is included and (ii) to
what extent theoretical uncertainties obtained in axial and triaxial
RHB calculations are correlated.

\begin{figure*}[ht]
\centering
\includegraphics[angle=0,width=8.0cm]{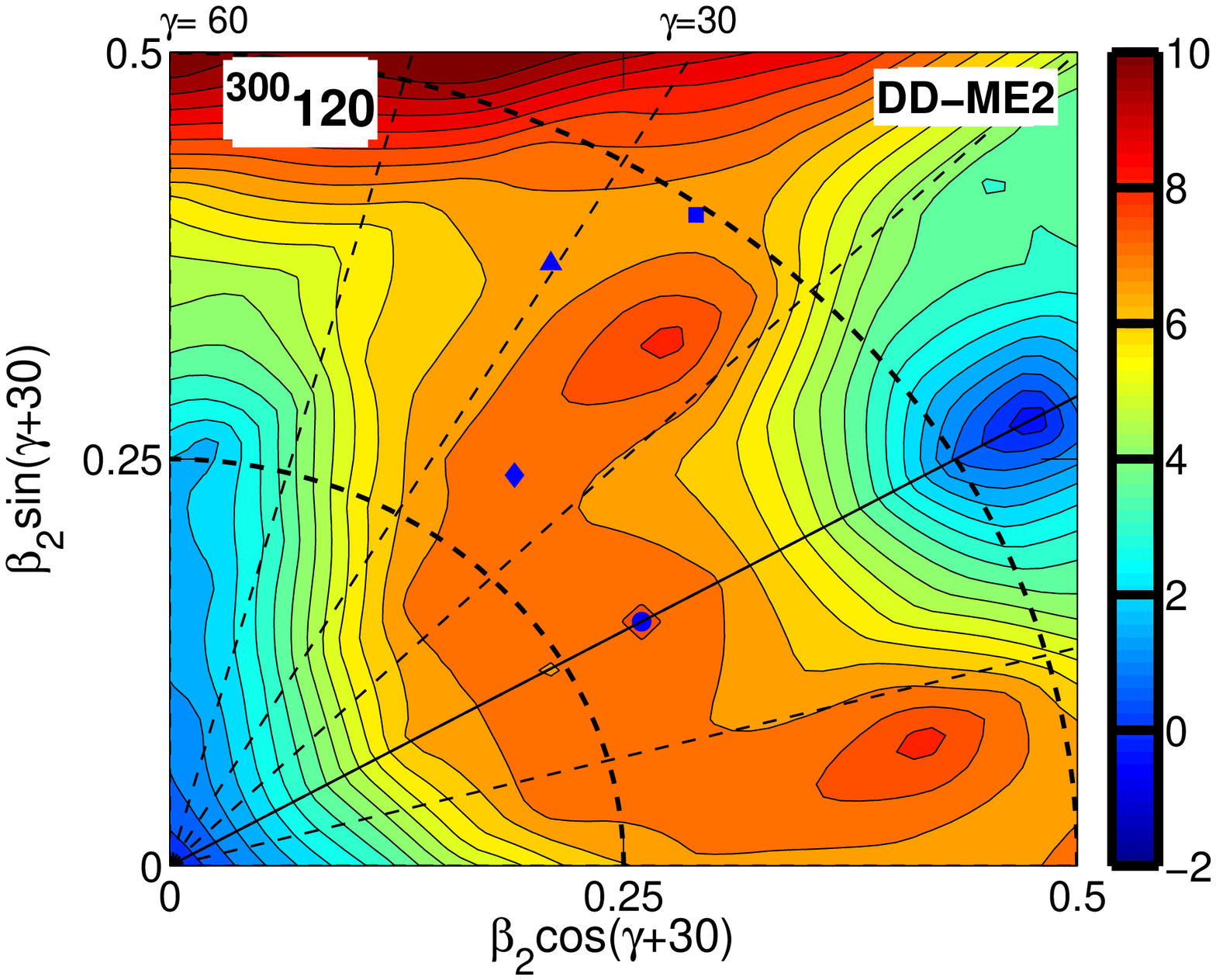}
\includegraphics[angle=0,width=8.0cm]{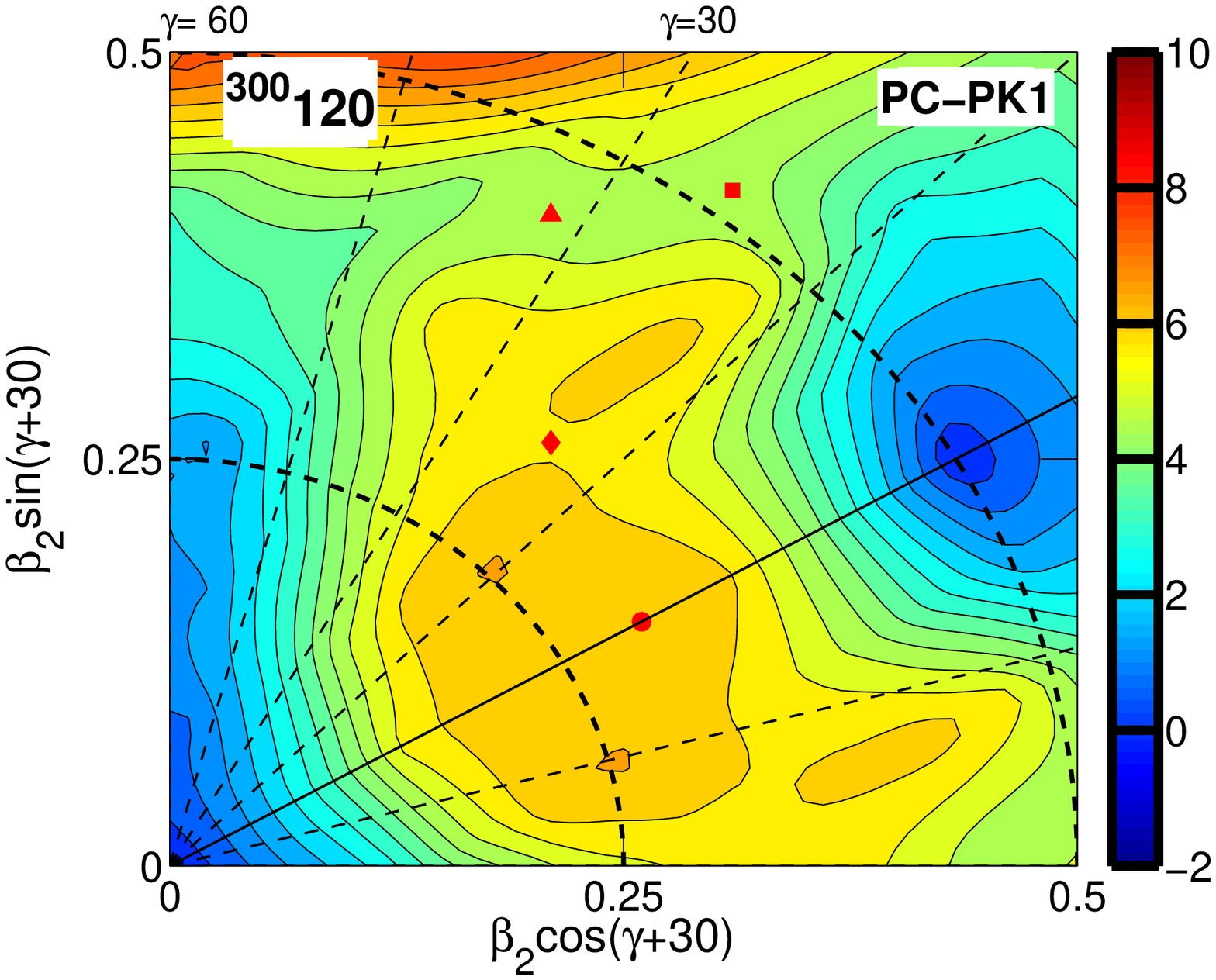}
\includegraphics[angle=0,width=8.0cm]{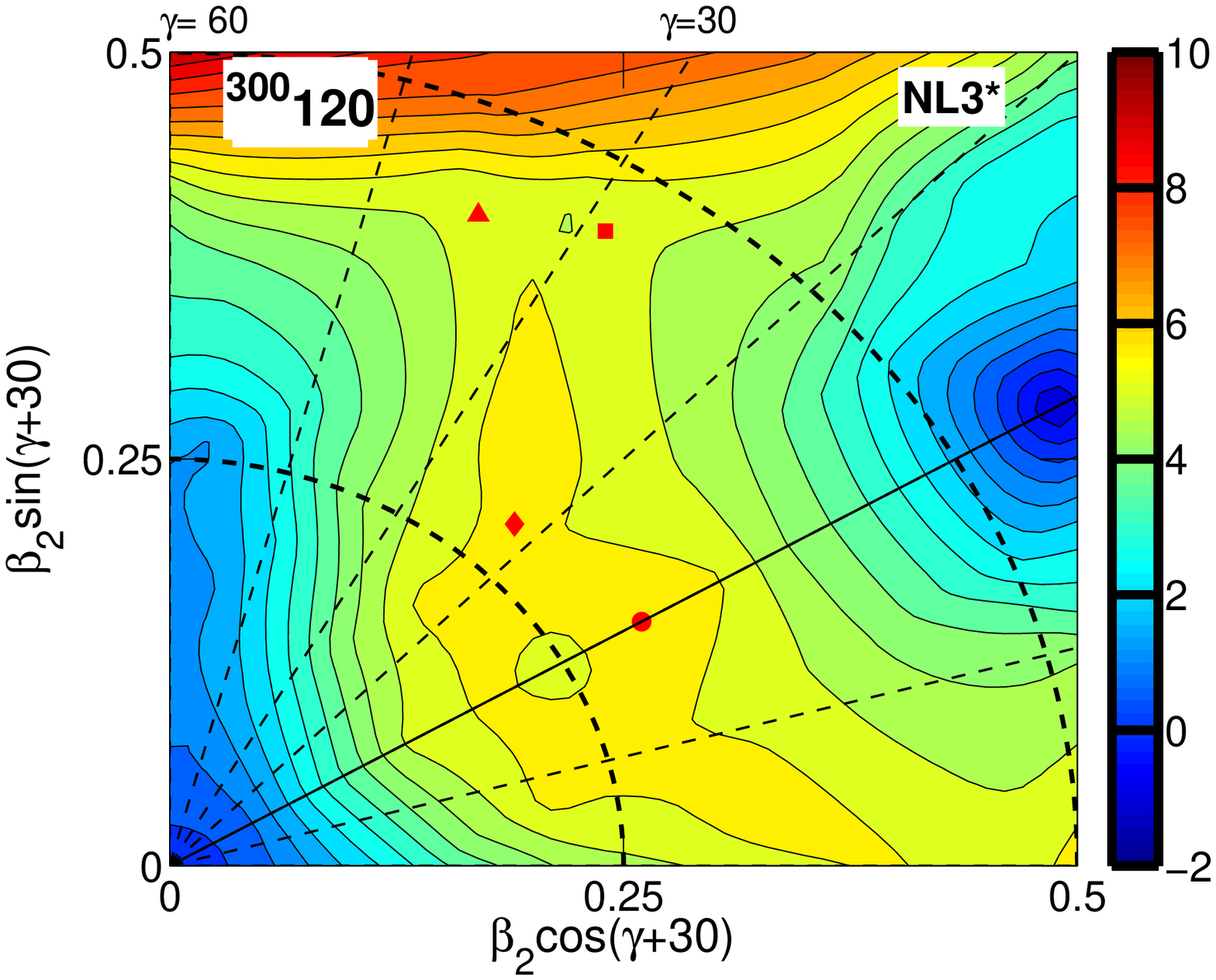}
\includegraphics[angle=0,width=8.0cm]{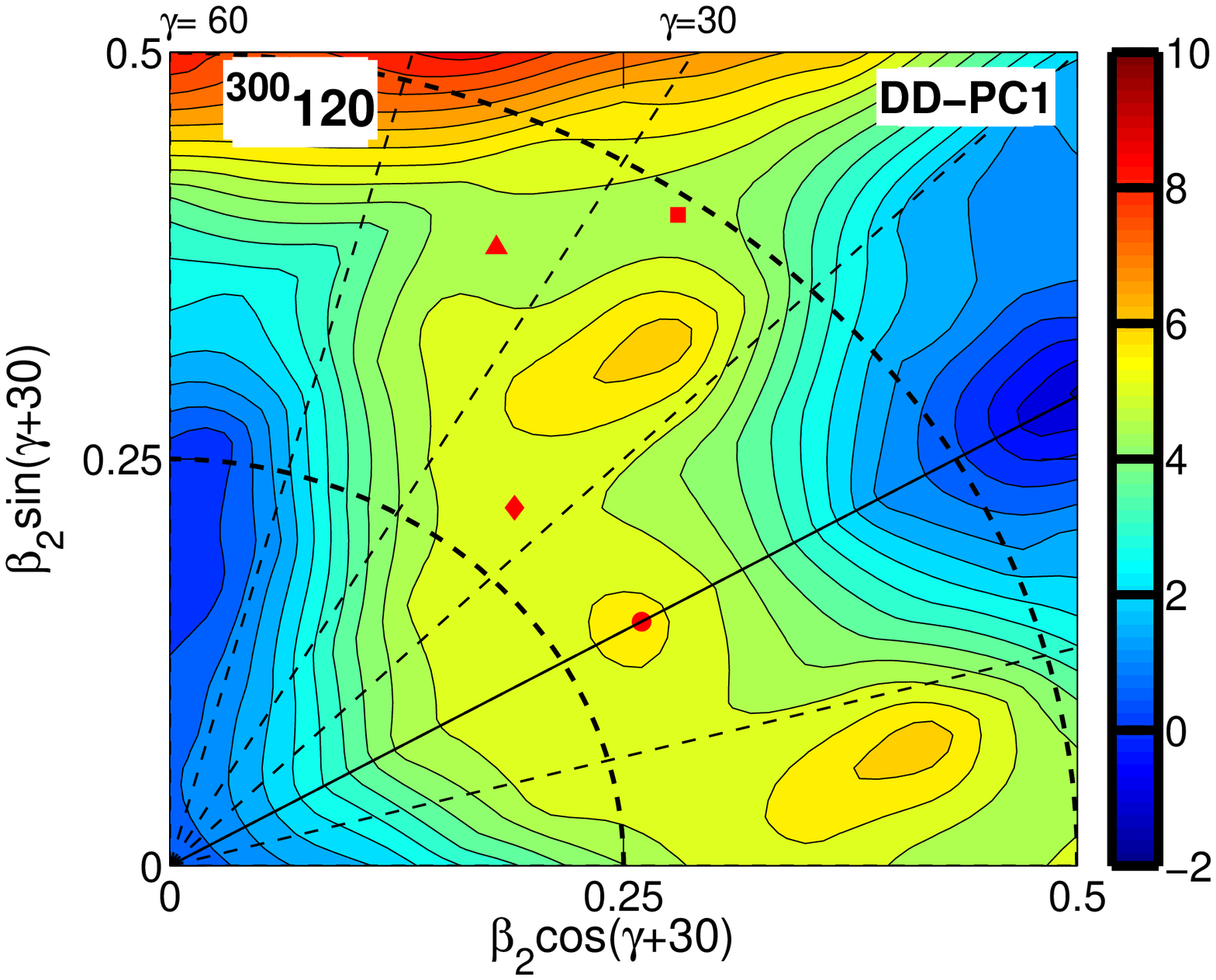}
\includegraphics[angle=0,width=8.0cm]{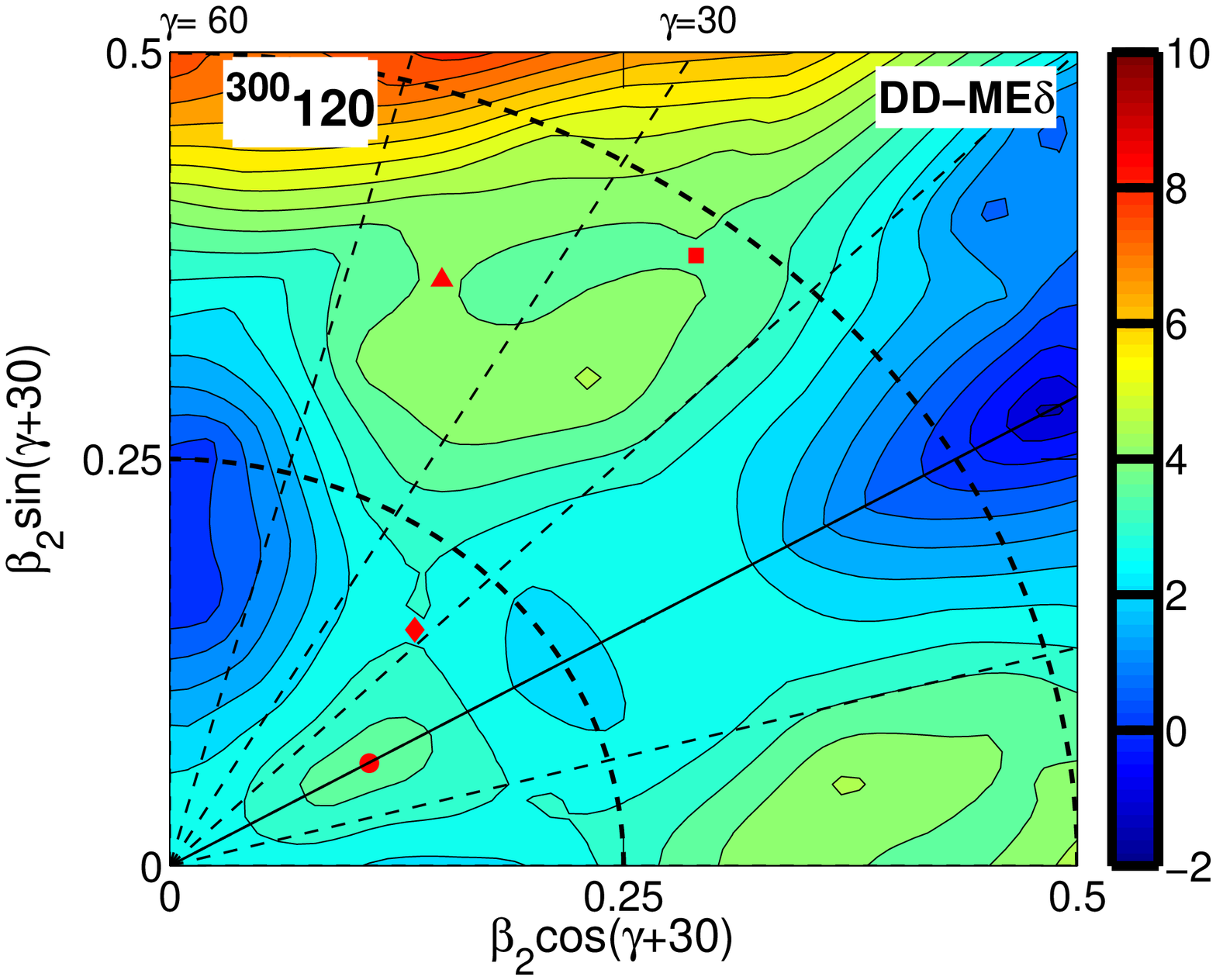}
\caption{(Color online) Potential energy surfaces of the $^{300}120$
nucleus as obtained in the calculations with indicated CEDFs. The
energy difference between two neighboring equipotential lines is
equal to 0.5 MeV. The Ax, Ax-Tr, Tr-A and Tr-B saddles are shown
by blue/red circles, diamonds, triangles, and squares, respectively.
The PES are shown in the order of decreasing height of inner
fission barrier.}
\label{PES-120-300}
\end{figure*}

\begin{figure*}[ht]
\centering
\includegraphics[angle=0,width=8.0cm]{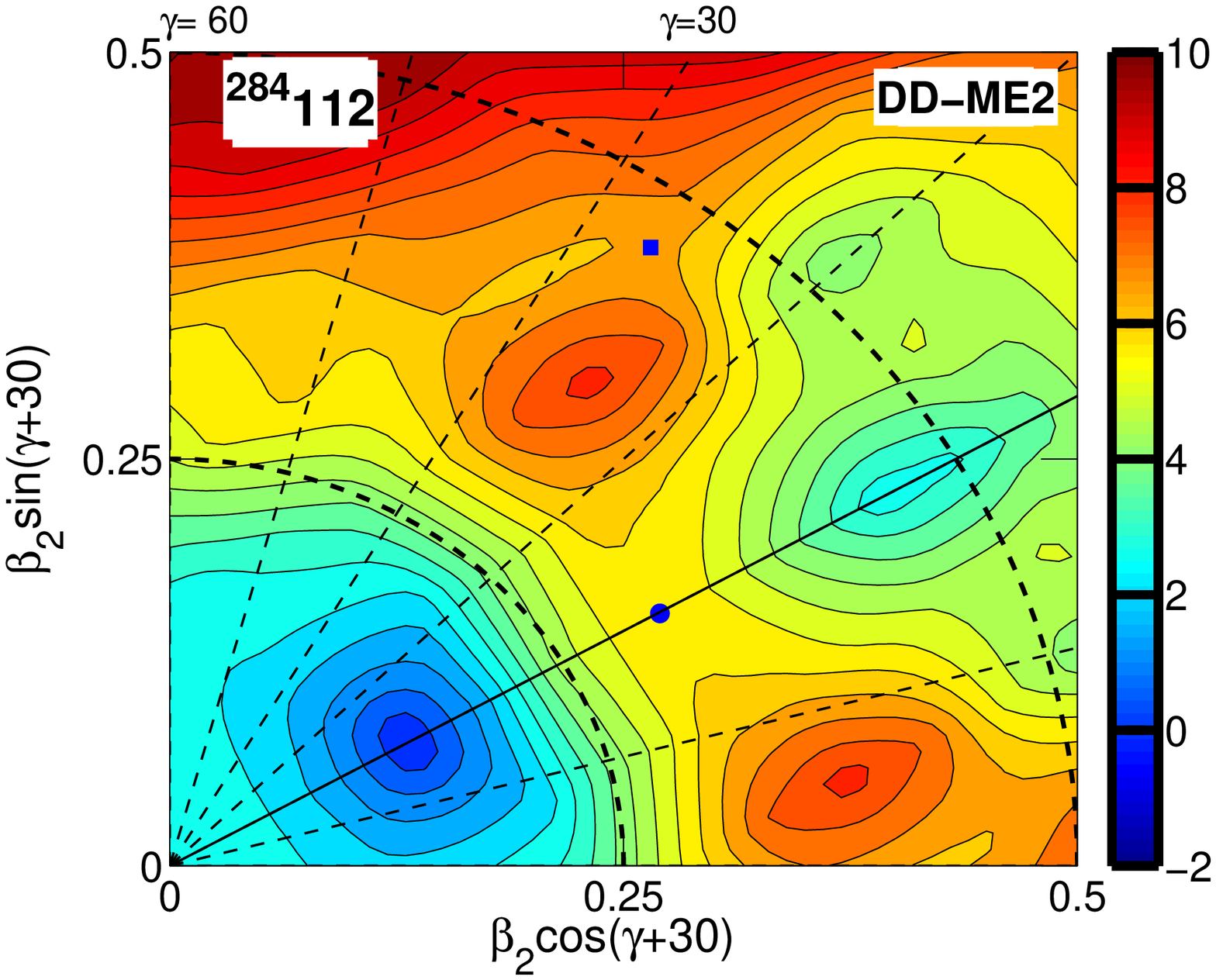}
\includegraphics[angle=0,width=8.0cm]{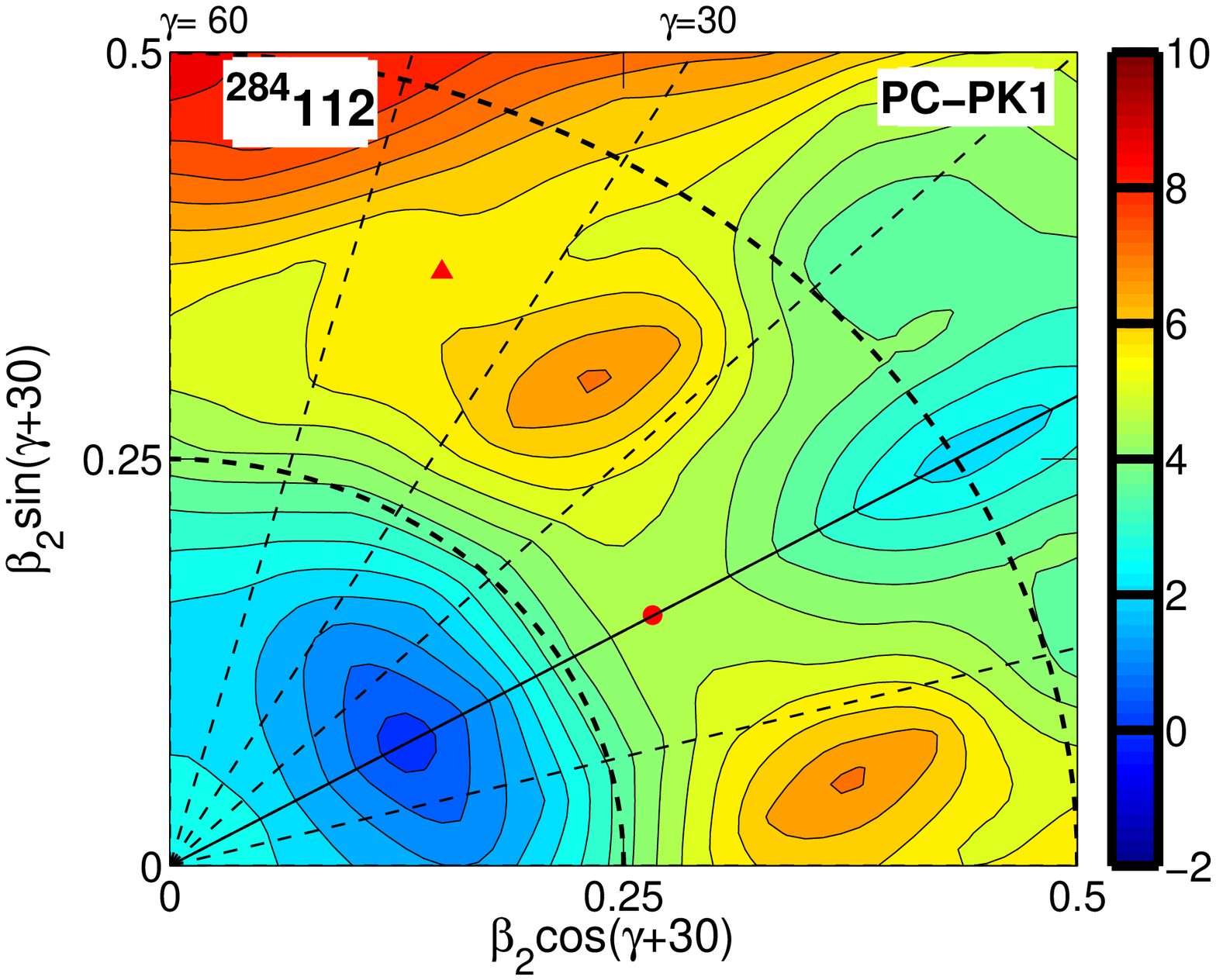}
\includegraphics[angle=0,width=8.0cm]{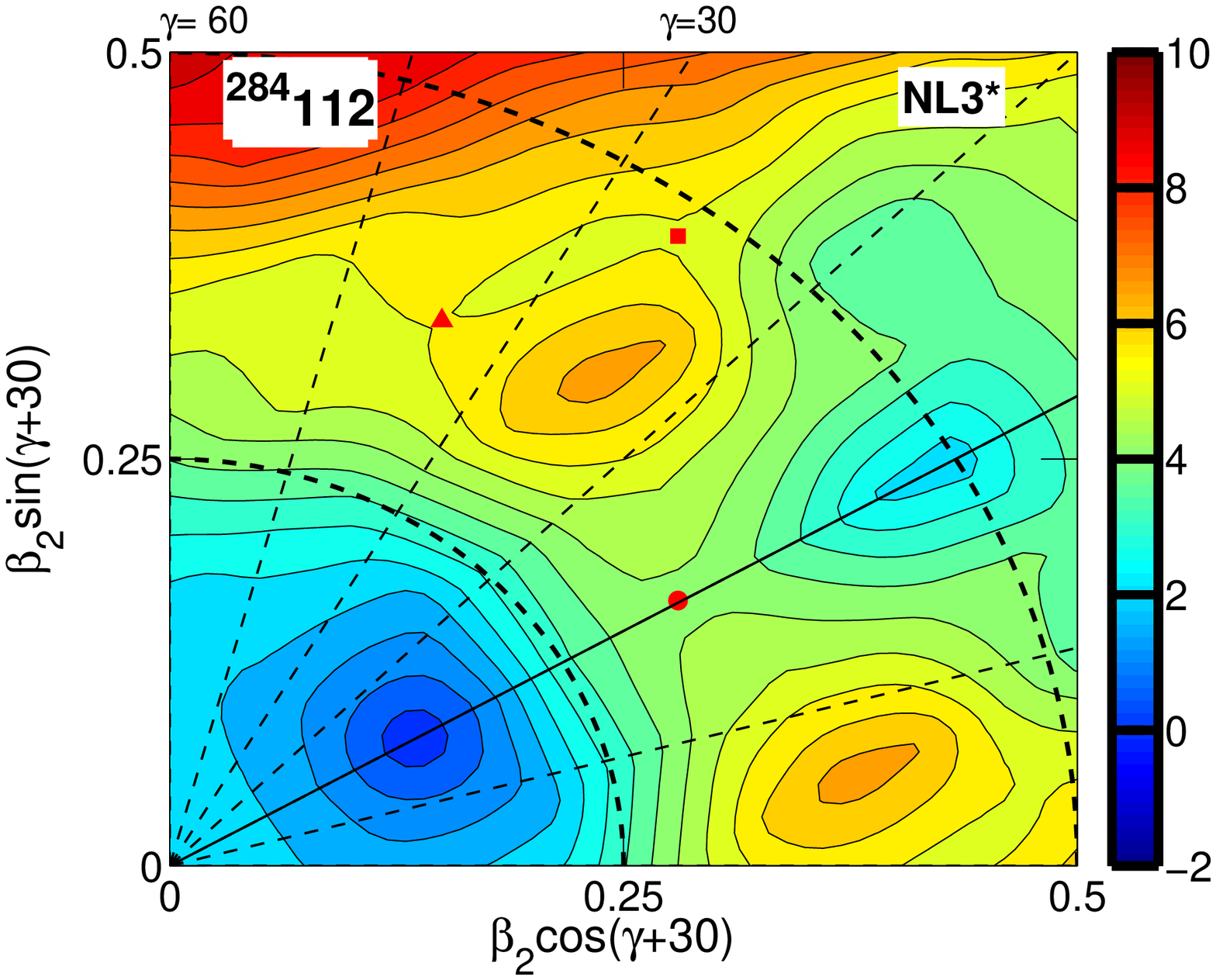}
\includegraphics[angle=0,width=8.0cm]{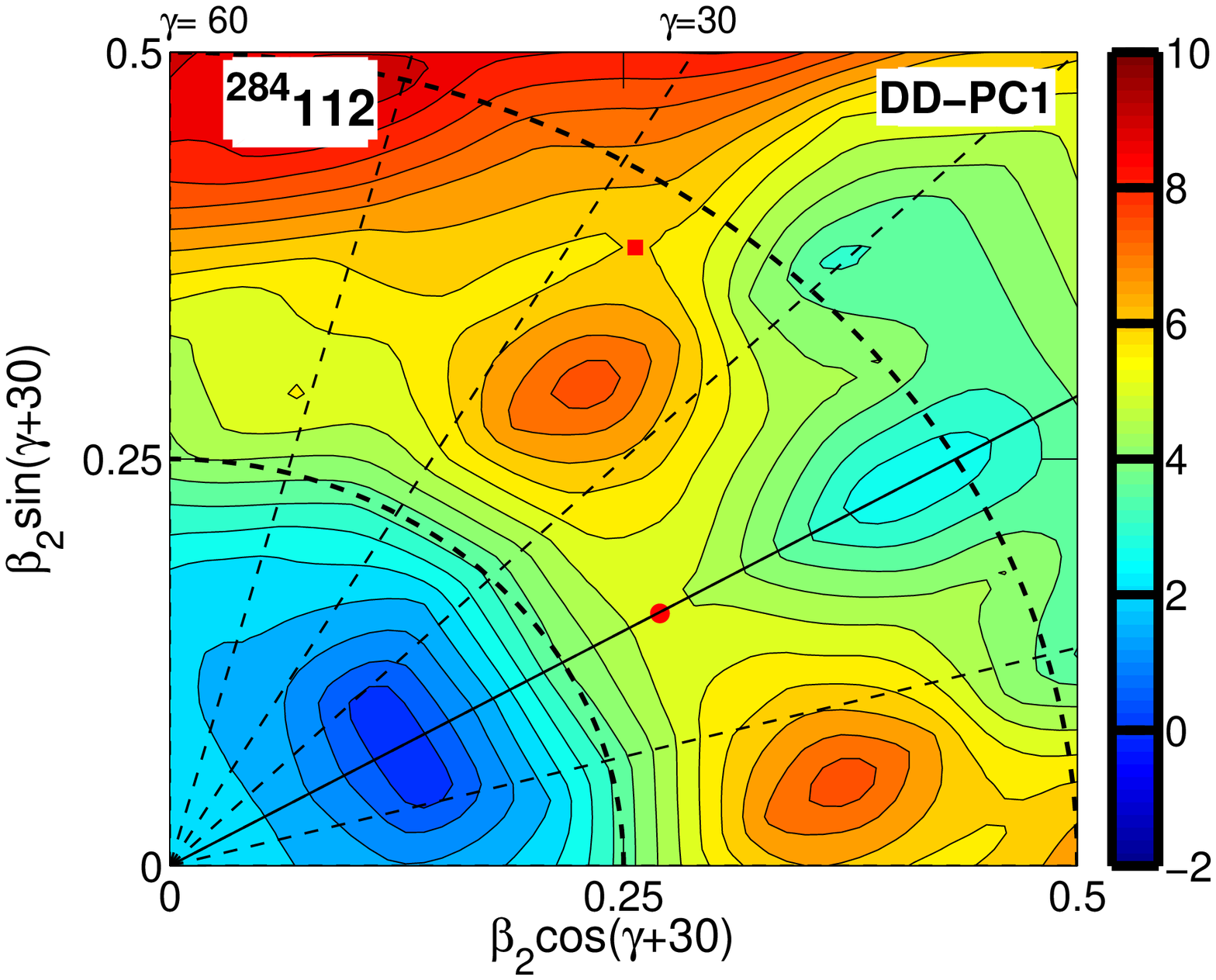}
\includegraphics[angle=0,width=8.0cm]{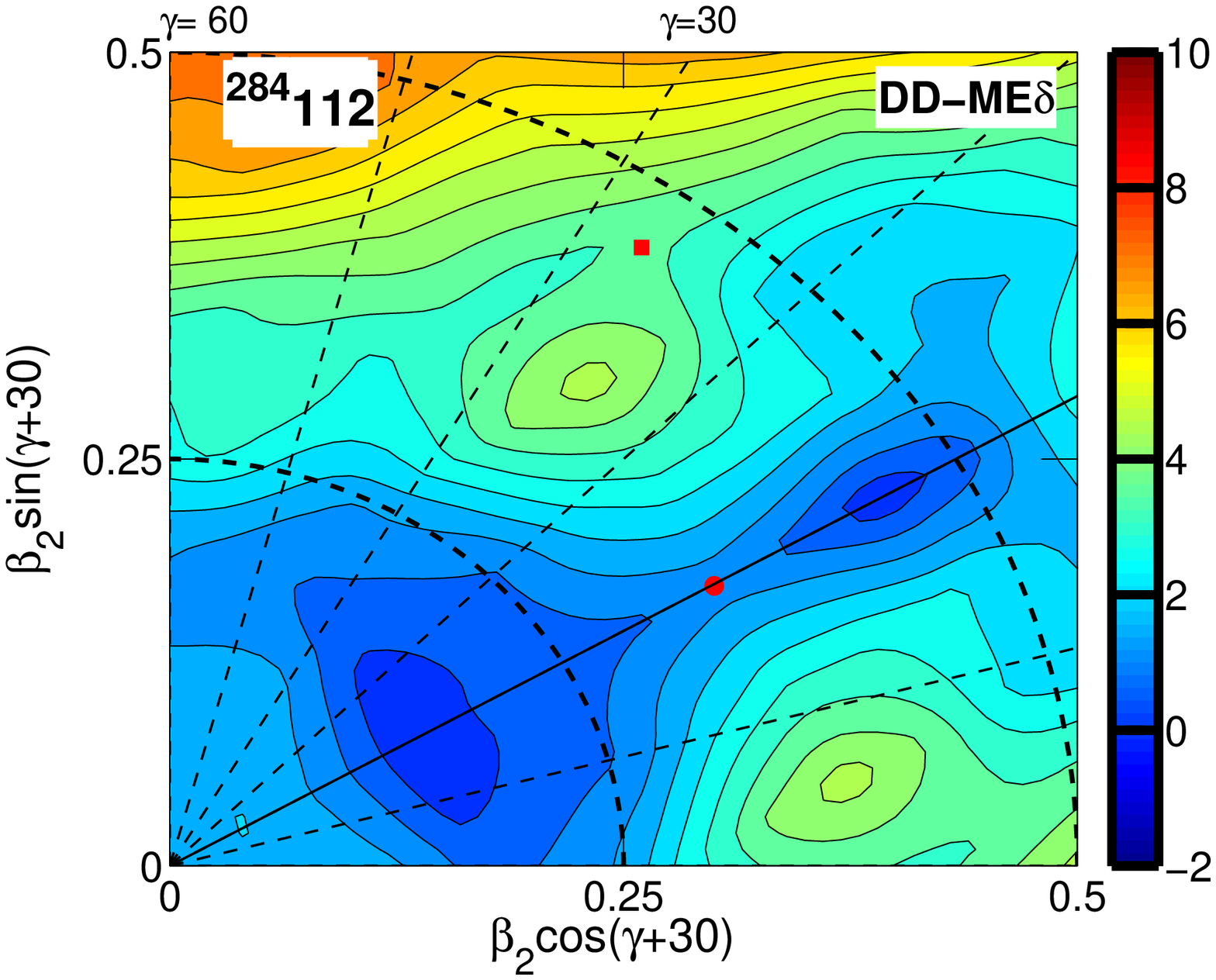}
\caption{(Color online) The same as in Fig.\ \ref{PES-120-300}
but for the $^{284}112$ nucleus.}
\label{PES-112-184}
\end{figure*}

  The dependence of the potential energy surfaces on the CEDF
is illustrated in Figs.\ \ref{PES-120-300} and \ref{PES-112-184}.
These PES are characterized by a complicated topology which, however,
reveals some typical triaxial saddles.

  For example, in the nucleus $^{300}$120 they are located at
$(\beta_2 \sim 0.32, \gamma \sim 21^{\circ})$, $(\beta_2 \sim 0.43,
\gamma \sim 33^{\circ})$, and $(\beta_2 \sim 0.49, \gamma \sim 24^{\circ})$
for the functionals DD-ME2, PC-PK1, NL3* and DD-PC1 (see Fig.\
\ref{PES-120-300}). The later two are also visible
in the CEDF DD-ME$\delta$. However, the first one is shifted
to smaller $\beta_2$ and $\gamma$ deformations, namely, to
$(\beta_2 \sim 0.20, \gamma \sim 15^{\circ})$.

 For all functionals except DD-ME$\delta$ the axial saddle is higher in energy
by roughly 0.5 MeV than the triaxial saddle at $(\beta_2 \sim 0.32, \gamma \sim 21^{\circ})$
and by approximately 1.5 MeV than the triaxial saddles at $(\beta_2 \sim 0.43, \gamma \sim 33^{\circ})$
and $(\beta_2 \sim 0.49, \gamma \sim 24^{\circ})$ (Fig.\ \ref{PES-120-300}).
The PES of the DD-ME$\delta$ functional has a completely different topology. Although
the  $(\beta_2 \sim 0.20, \gamma \sim 15^{\circ})$ saddle is lower in energy than the
axial saddle by approximately 1 MeV, the axial saddle is located only $\sim 0.2$ MeV
below the triaxial saddles at $(\beta_2 \sim 0.33, \gamma \sim 25^{\circ})$ and
$(\beta_2 \sim 0.45, \gamma \sim 33^{\circ})$.

 The presence of these saddles leads to several fission paths which have
been discussed in detail in Ref.\ \cite{AAR.12}.  Although this discussion
is based on RMF+BCS results with NL3*, we found that it is still valid for
the TRHB results with DD-ME2, PC-PK1, NL3* and DD-PC1. This is because for
a given functional the topology of PES obtained in triaxial RMF+BCS and RHB
calculations is similar. The situation is different for DD-ME$\delta$ which
has an axial saddle located at $\beta_2 \sim 0.13$ (Fig.\ \ref{PES-120-300}).
Thus, the fission path will proceed from the oblate minimum via the triaxial
saddle at $(\beta_2 \sim 0.20, \gamma \sim 0.15)$ which has a low excitation
energy of only 3 MeV.

  As shown in Ref.\ \cite{AAR.12}, the axial saddle becomes energetically
more favored as compared with triaxial saddles on moving away from the
particle numbers $Z=120$ and $N=184$. This is clearly seen in the nucleus
$^{284}$112, in which the axial saddle at $\beta_2 \sim 0.32$ is lower in
energy than the triaxial saddles located around $(\beta_2 \sim 0.38, \gamma \sim 34^{\circ})$
and $(\beta_2 \sim 0.47, \gamma \sim 26^{\circ})$. This feature is also seen in
Fig.\ 4 of Ref.\ \cite{AAR.12} which compares the results for selected $Z=112, 114$,
and 116 nuclei obtained in the RMF+BCS calculations with NL3*.

  To simplify the further discussion we follow the notation of Ref.\
\cite{AAR.12} and denote the axial saddle as 'Ax', the triaxial saddle
with  ($\beta_2 \sim 0.3, \gamma \sim 10^{\circ}$) as 'Ax-Tr', the
triaxial ($\beta_2 \sim 0.4, \gamma \sim 35^{\circ}$) saddle as 'Tr-A'
and the triaxial saddle with ($\beta_2 \sim 0.5, \gamma \sim 22^{\circ}$)
as 'Tr-B'. Although the positions of these saddles move somewhat
in the deformation plane with the change of proton and neutron numbers,
they appear in the majority of nuclei under study.

\begin{figure*}[ht]
\centering
\includegraphics[angle=0,width=14.0cm]{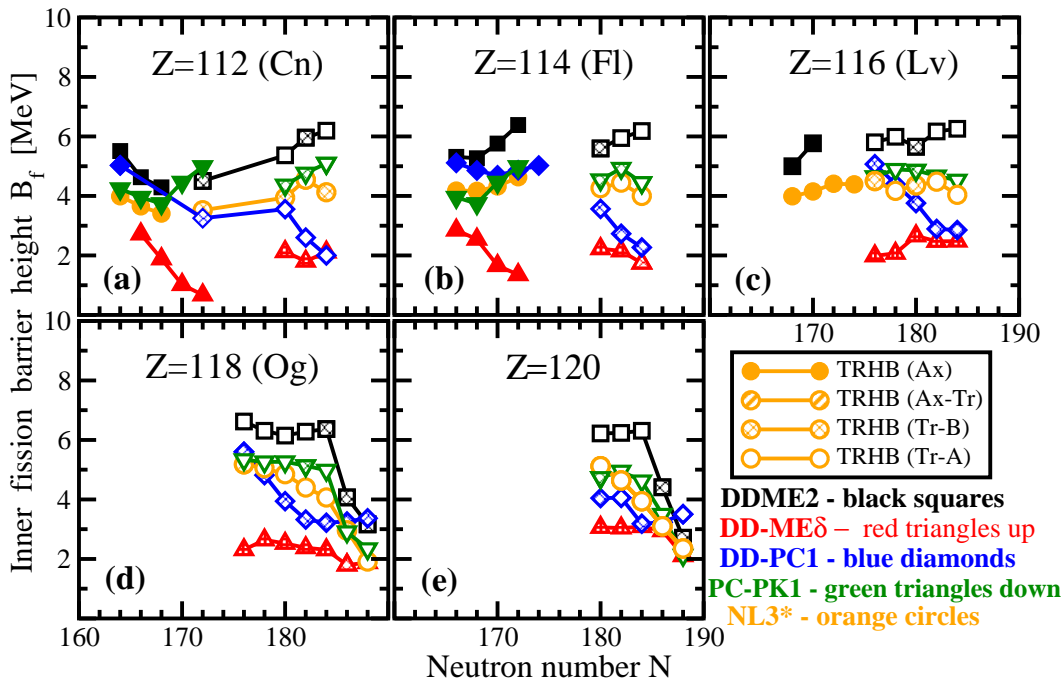}
\caption{(Color online) The heights of inner fission barriers in
selected nuclei as obtained in the TRHB calculations with indicated
CEDFs. The style of the symbol filling indicates the type of the
lowest in energy saddle. Note that the TRHB results in a few
$N\sim 166$ and $N=172$ nuclei (see Fig.\ \ref{fission_spread_triax}
below) and the trends of the evolution of PES with
particle number allow to firmly establish the axial symmetric nature of
the lowest saddle in the $Z=112$ and 114 nuclei (as well as in $Z=116$
nuclei for the NL3* and DD-ME2 functionals) for neutron numbers
between 164 and approximately $N=172$. For some of these nuclei,
we use axial RHB results when the TRHB results are not available.}
\label{FB-SHE-triaxial}
\end{figure*}

  Fig.\ \ref{FB-SHE-triaxial} summarizes the results of the calculations
for the inner fission barrier heights. The DD-ME2 and DD-ME$\delta$
functionals provide the highest and the lowest fission barriers among
those obtained in the calculations with five CEDFs.  The results of the
calculations with the CEDFs NL3*, DD-PC1 and PC-PK1 are located in between
of these two extremes. Note that these three functionals have been benchmarked
in the actinides in Refs.\ \cite{AAR.12,LZZ.12,PNLV.12,LZZ.14} where they provide
a good description of experimental data.

  Fig.\ \ref{FB-SHE-triaxial} clearly shows that different functionals are
characterized by different isotopic and isotonic dependencies for the inner
fission barrier heights. As a result, the functionals, which give similar
results in one part of the $(Z,N)$ plane, could provide substantially different
results in another. This leads to the spreads in the predictions of the
inner fission barrier heights which are presented in Figs.\ \ref{fission_spread_triax}
and \ref{fission_spread_triax_1}. The strongest
correlation between these spreads is observed for the 'Ax' and 'Ax-Tr' saddles;
this is seen both for the set of five (Figs.\ \ref{fission_spread_triax}a and b)
and the set of three (Figs.\ \ref{fission_spread_triax_1}a and b) functionals.
This is because these saddles are closely located in the deformation plane so
that the change in the energy of the 'Ax' saddle affects in a similar way the
energy of the 'Ax-Tr' saddle.  The correlations in the spreads of the energies
of the 'Ax' saddle on one hand and the 'Tr-A' and 'Tr-B' saddles on the other
hand depends on how many functionals are used in the analysis. On average,
they are strongly correlated for the set of the DD-PC1, NL3* and PC-PK1
functionals (compare Figs. \ref{fission_spread_triax_1}a, c
and d) which have large similarities in the topology of PESs (see Figs.\
\ref{PES-120-300} and \ref{PES-112-184}) and for which the $\Delta E^S$
spreads are typically below 2 MeV (see Fig.\ \ref{fission_spread_triax_1}).
Note that these three functionals successfully describe experimental fission
barriers in the actinides \cite{AAR.10,PNLV.12,LZZ.12}. These correlations
decrease with the addition of the functionals DD-ME$\delta$ and DD-ME2; the
$\Delta E^s$ spreads are typically smaller for the 'Tr-A' and 'Tr-B' saddles
as compared with the 'Ax' one (compare Figs. \ref{fission_spread_triax}a,
c and d).

  It is important that the spreads for the axial 'Ax' saddles and the lowest in
energy saddles are strongly correlated (compare Figs.\ \ref{fission_spread_triax}a
and d and Figs.\ \ref{fission_spread_triax_1}a and d). This suggests that also for
other regions of nuclear chart, not covered by the present TRHB calculations,
the spreads in inner fission barrier heights obtained in the axial RHB calculations
could be used as a reasonable estimate of the spreads which would be obtained
in the calculations with triaxiality included.

\begin{figure*}[ht]
\centering
\includegraphics[angle=-90,width=8.8cm]{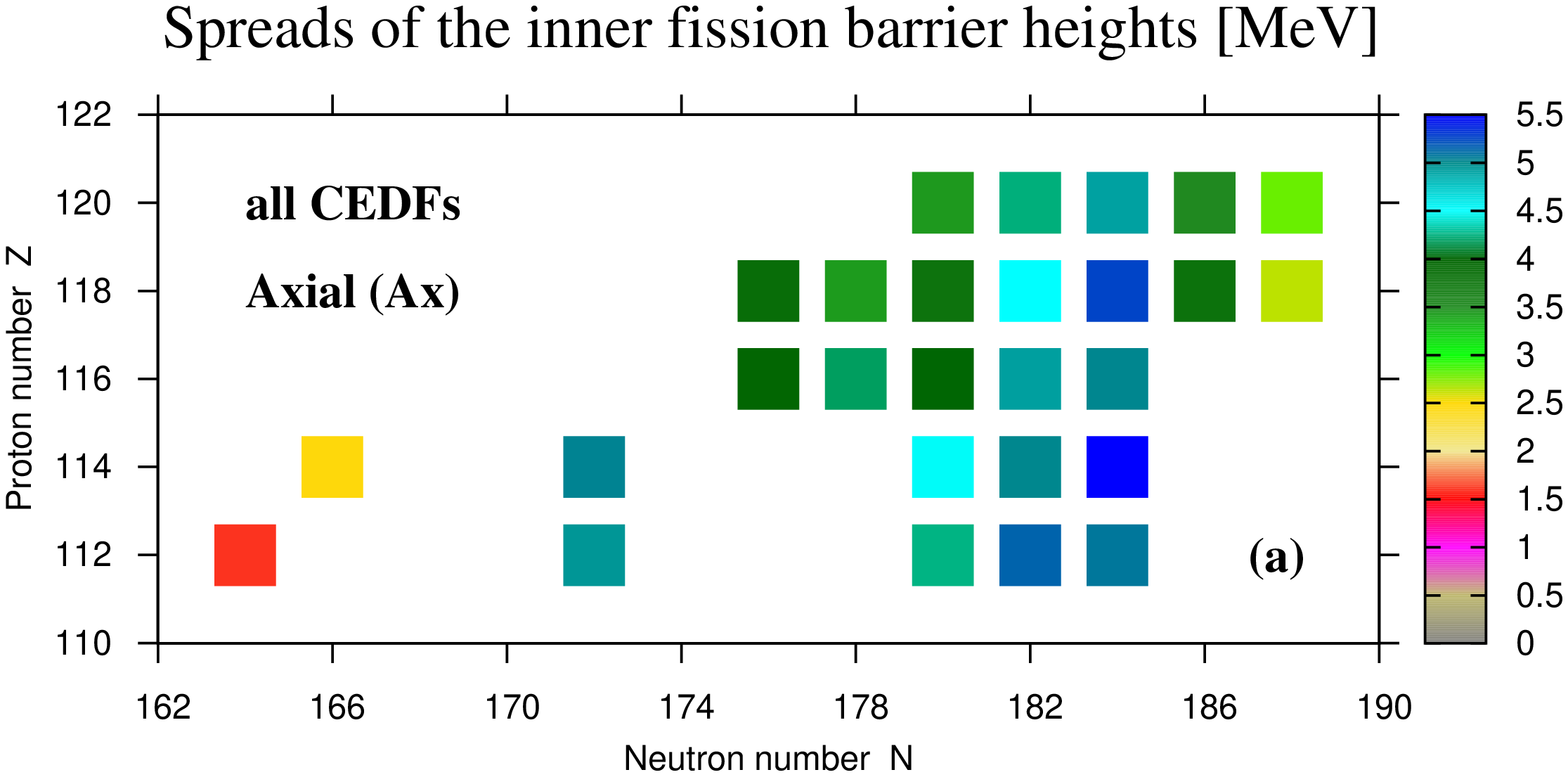}
\includegraphics[angle=-90,width=8.8cm]{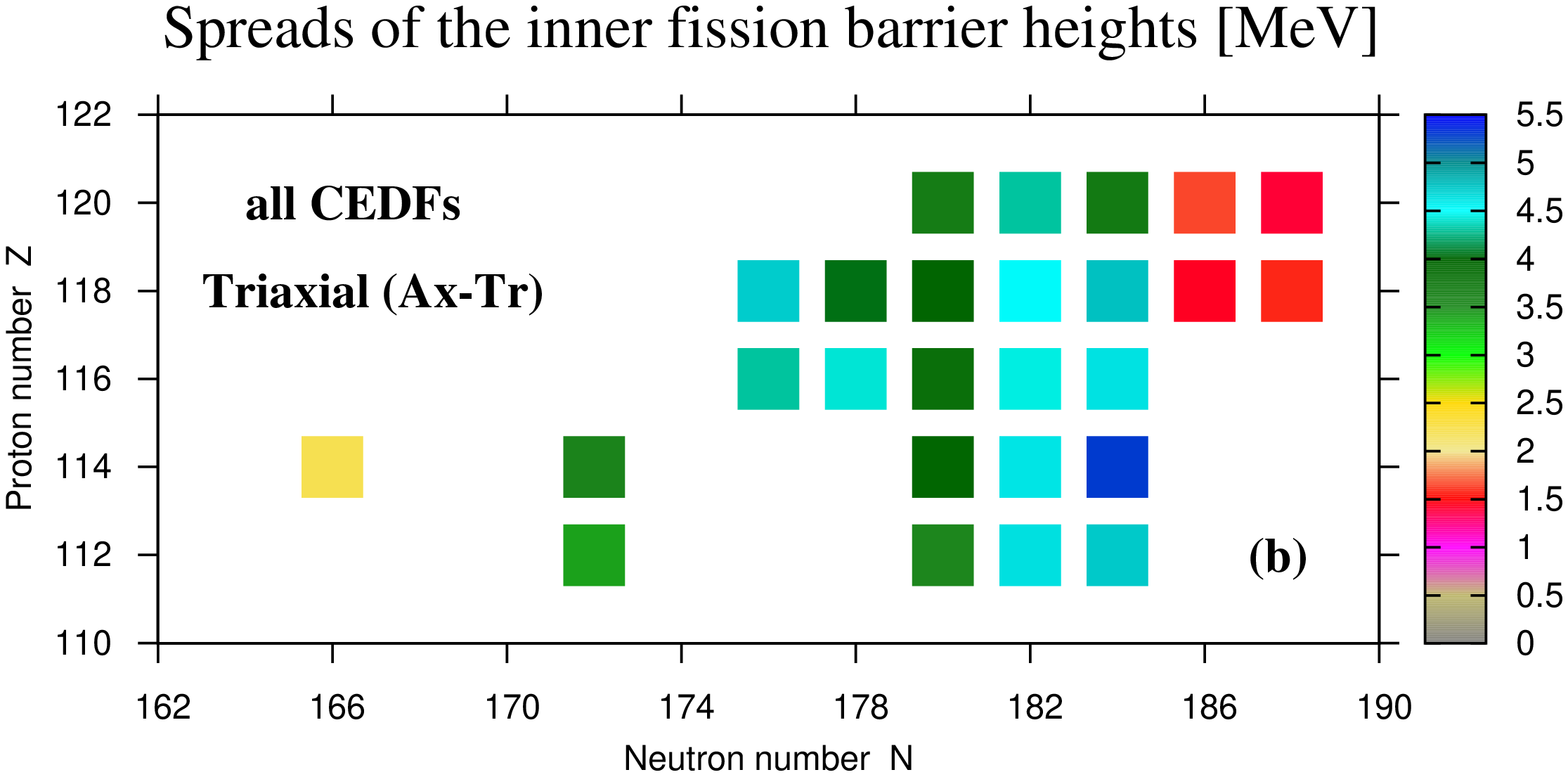}
\includegraphics[angle=-90,width=8.8cm]{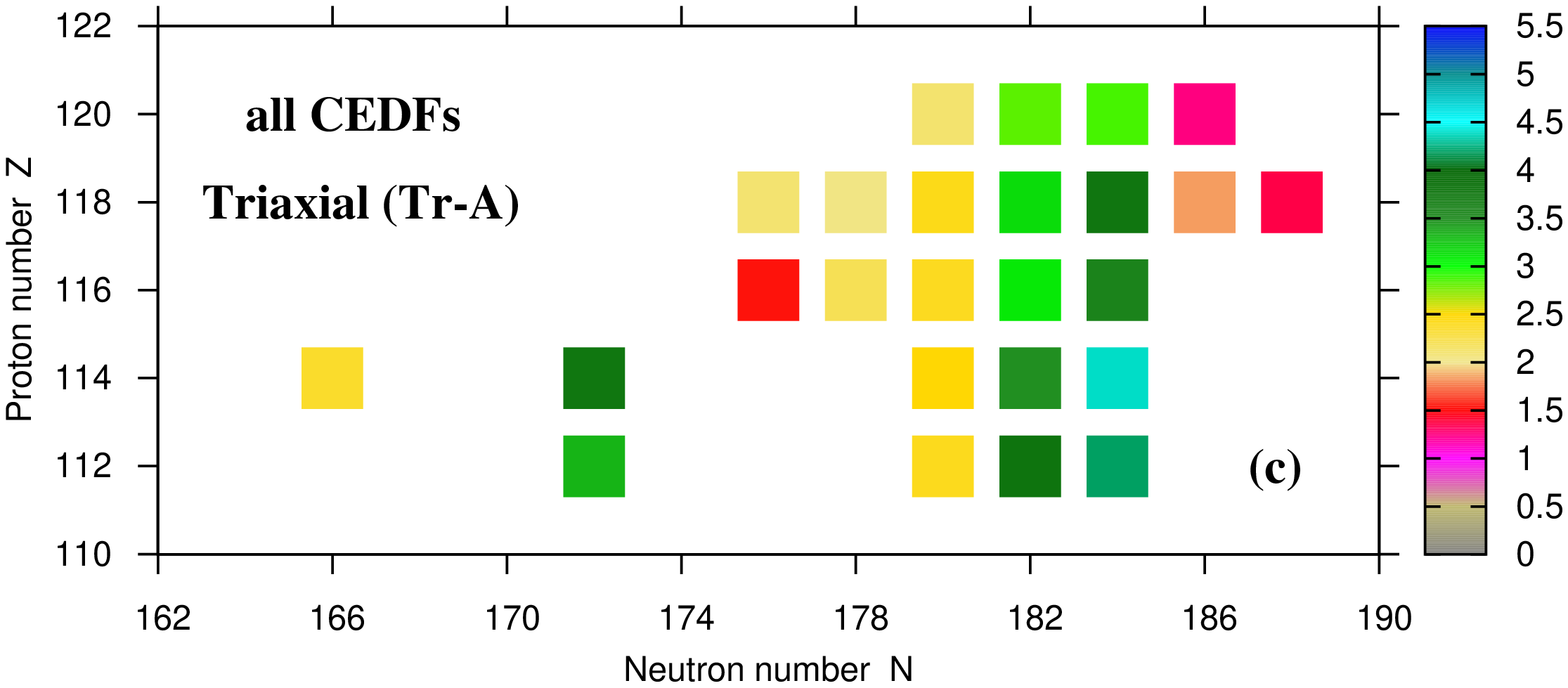}
\includegraphics[angle=-90,width=8.8cm]{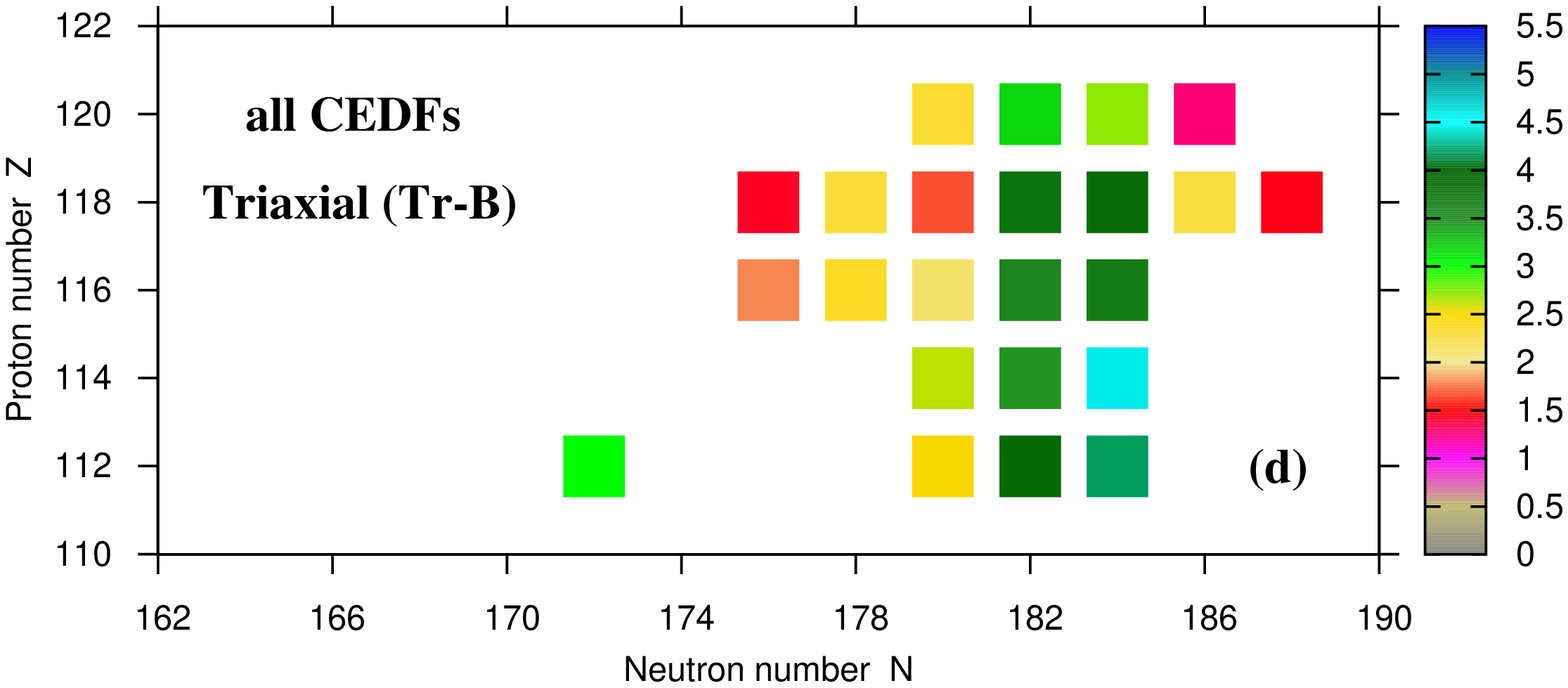}
\includegraphics[angle=-90,width=8.8cm]{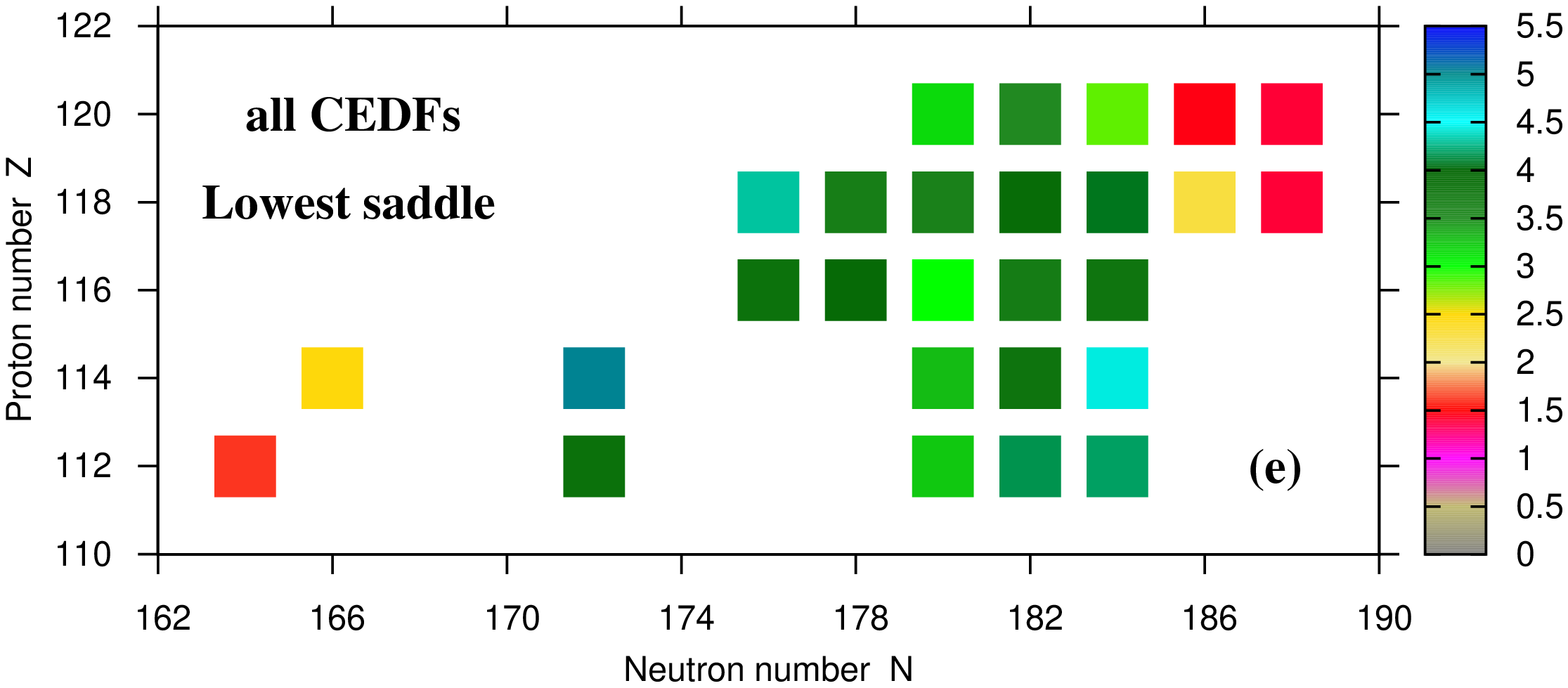}
\caption{(Color online) The spreads $\Delta E^S$ of the energies
of axial (panel (a)), triaxial (panels (b,c,d)) and the
lowest in energy (panel (d)) saddles for a selected set of the $Z=112-120$
nuclei as a function of proton and
neutron number. $\Delta E^S(Z,N) = |E^{S}_{max}(Z,N)-E^{S}_{min}(Z,N)|$, where, for
given $Z$ and $N$ values, $E^{S}_{max}(Z,N)$ and $E^{S}_{min}(Z,N)$ are the largest
and smallest energies of the saddles obtained with the set of functionals NL3*,
DD-ME2, DD-ME$\delta$, DD-PC1, and PC-PK1. Note that the same colormap
as in Fig.\ \ref{fission_spread} is used here.}
\label{fission_spread_triax}
\end{figure*}

\begin{figure*}[ht]
\centering
\includegraphics[angle=-90,width=8.8cm]{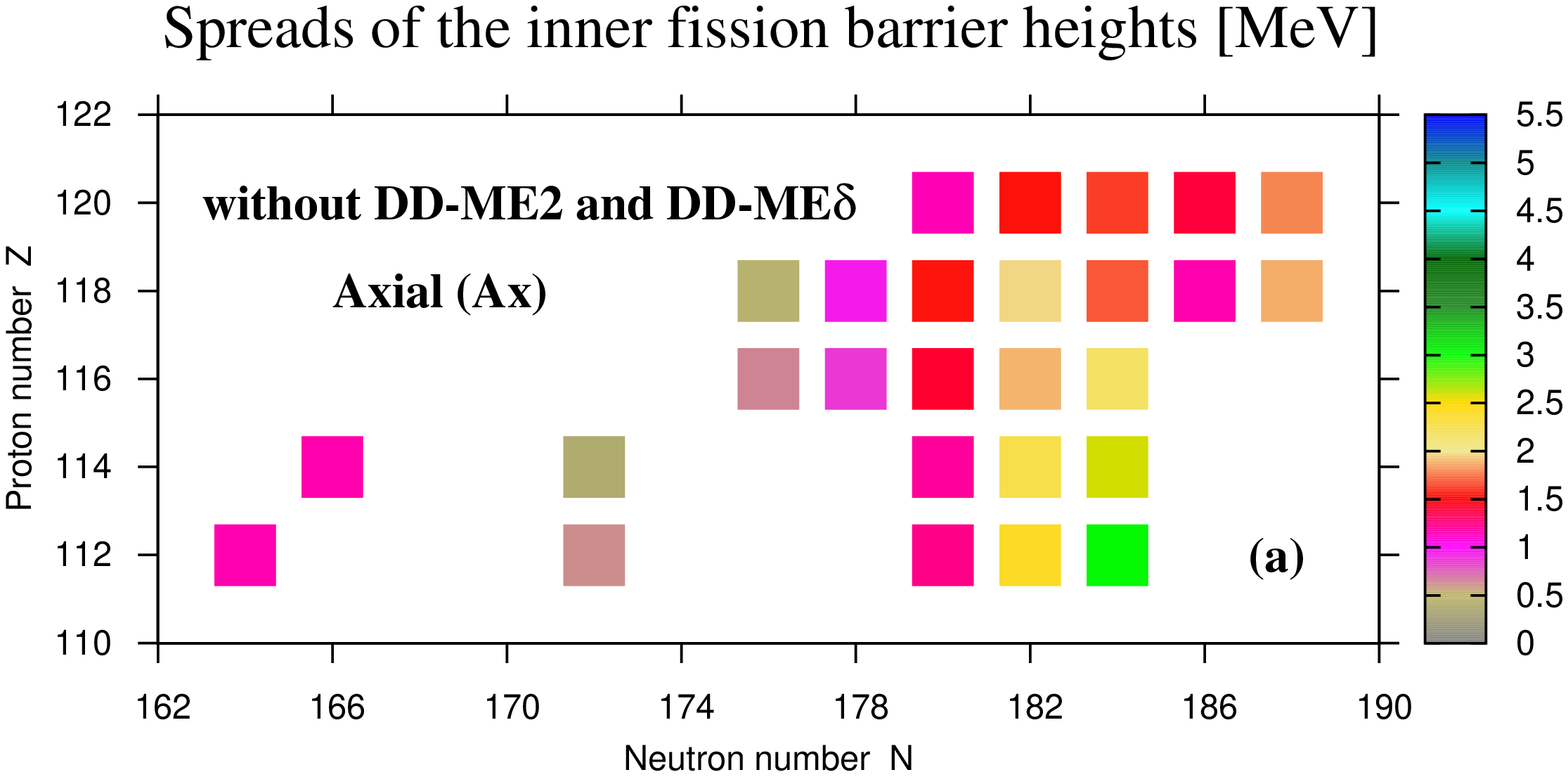}
\includegraphics[angle=-90,width=8.8cm]{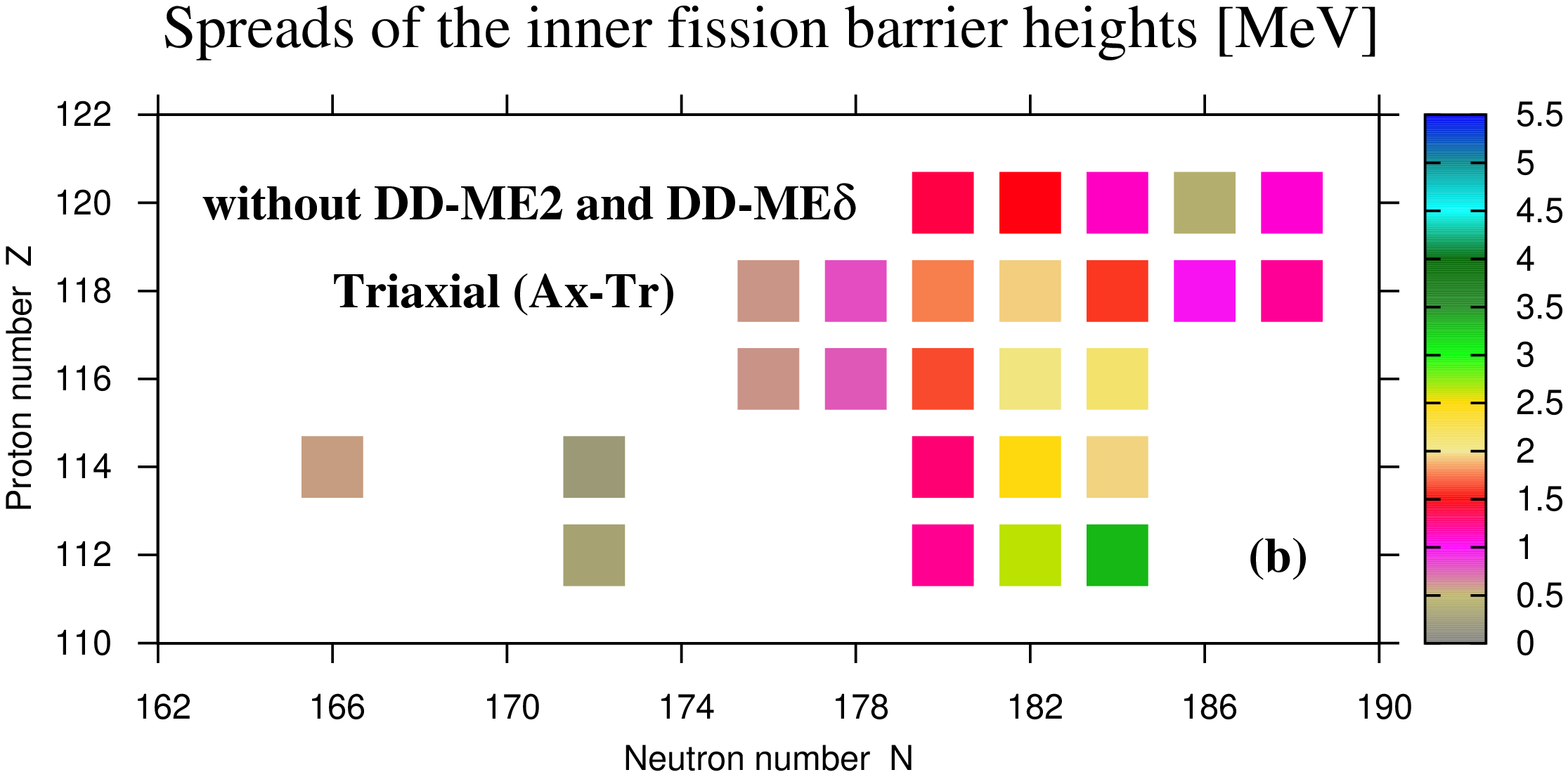}
\includegraphics[angle=-90,width=8.8cm]{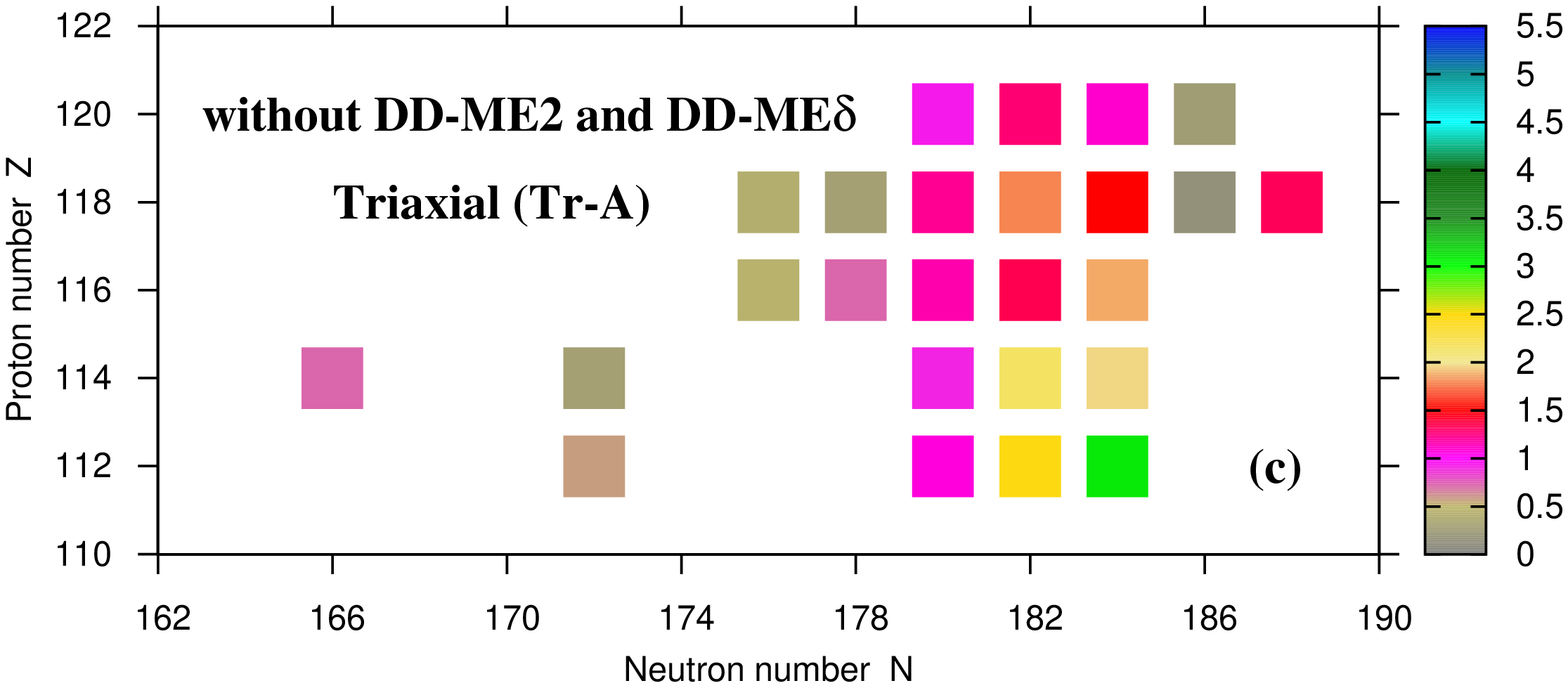}
\includegraphics[angle=-90,width=8.8cm]{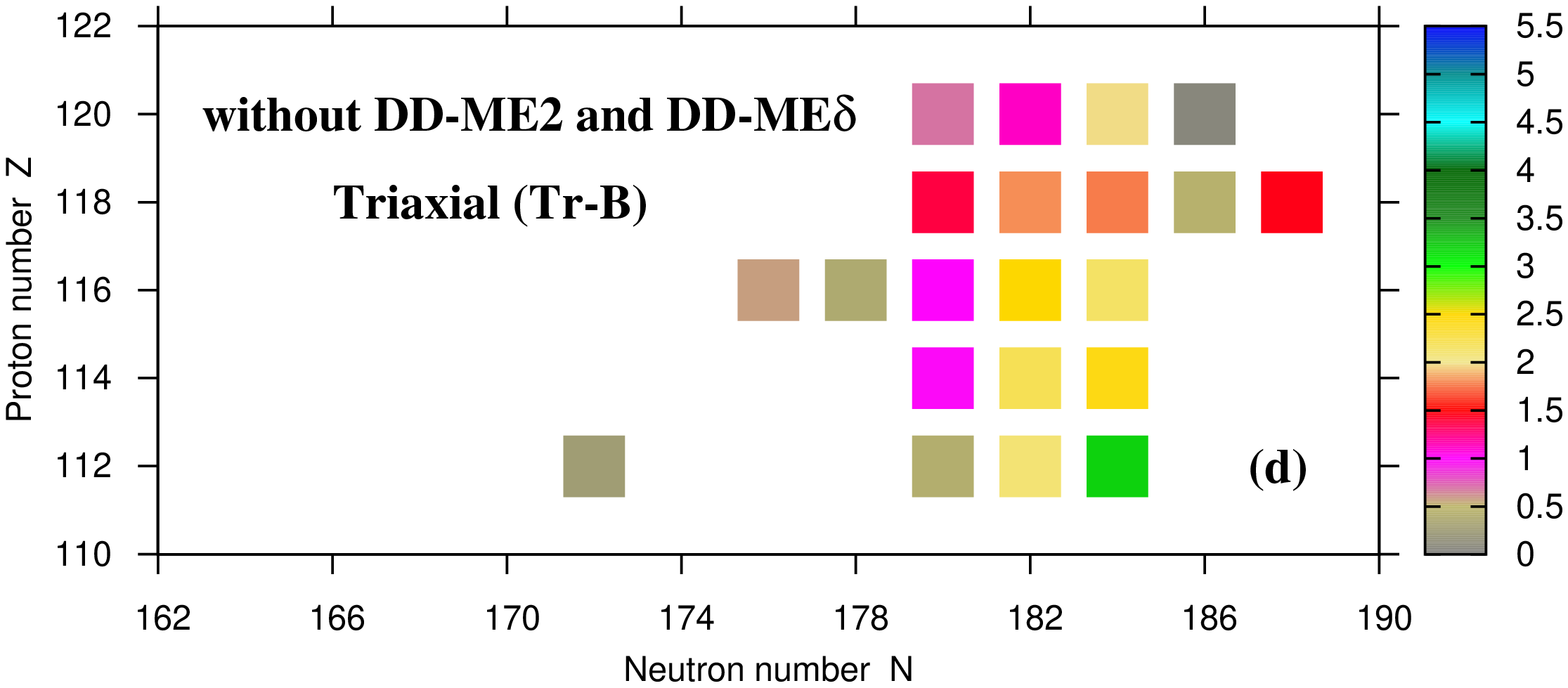}
\includegraphics[angle=-90,width=8.8cm]{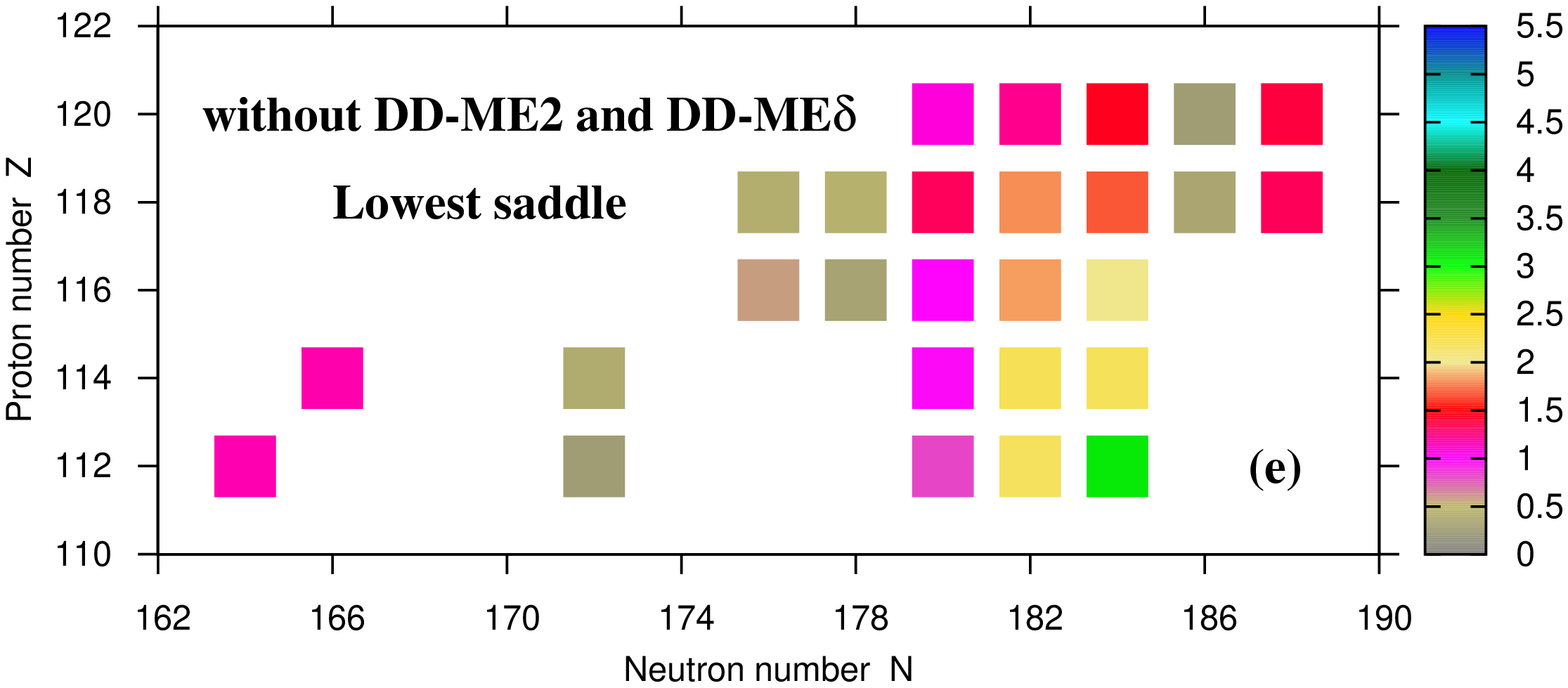}
\caption{(Color online) The same as in Fig.\ \ref{fission_spread_triax} but for
the case when the DD-ME2 and DD-ME$\delta$ CEDFs are excluded from
consideration.}
\label{fission_spread_triax_1}
\end{figure*}

\section{Comparison of fission barriers in different models}
\label{fission-barrier-dif-mod}

\begin{figure*}[ht]
\centering
\includegraphics[angle=0,width=14.0cm]{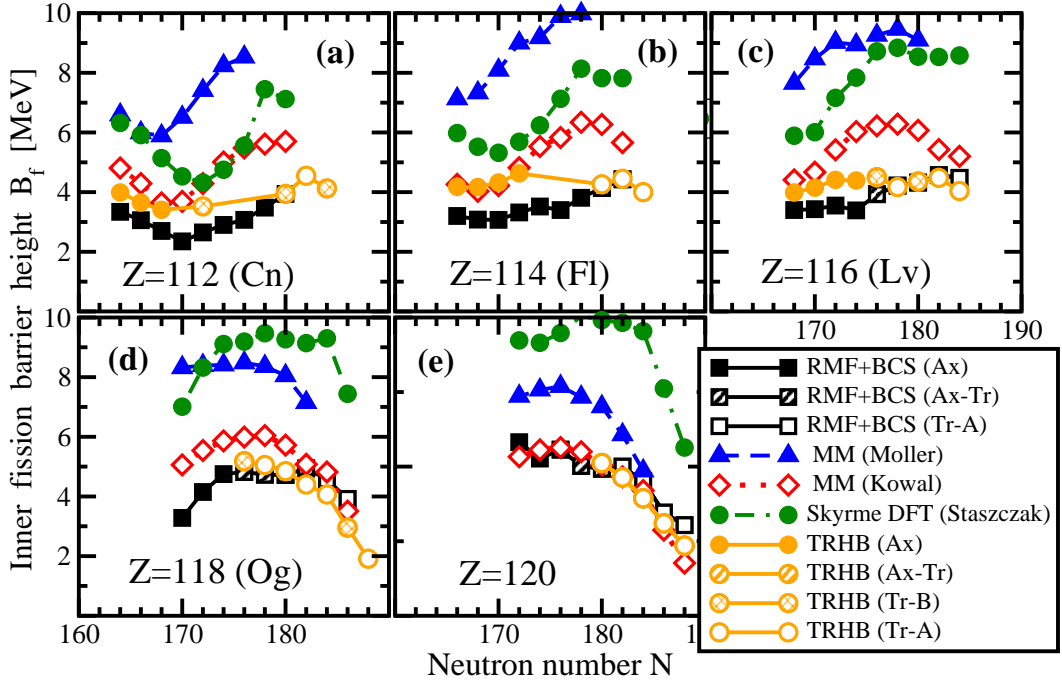}
\caption{Inner fission barrier heights $B_f$ as a function of
the neutron number $N$.  The position of the inner fission barrier
saddle in deformation space varies as a function of particle number.
Thus, the lowest saddles are labelled by 'Ax', 'Ax-Tr', 'Tr-A' and
'Tr-B' (see text for details). The results of triaxial RMF+BCS
calculations are taken from Ref.\ \cite{AAR.12}. The results of Skyrme
DFT calculations with SkM* have been taken from Ref.\ \cite{SBN.13}.
The results of the MM calculations are taken from Ref.\ \cite{MSI.09}
(labeled as 'MM (M{\"o}ller)') and Ref.\ \cite{KJS.10} (labeled as
'MM (Kowal)'). Note that the style of Fig.\ \ref{FB-SHE-axial} is used
here for easy comparison between two figures.
}
\label{FB-SHE}
\end{figure*}

   It is necessary to recognize that the CDFT represents only one of
the classes of nuclear structure models. Other classes are represented
by non-relativistic DFTs based on zero-range Skyrme and finite range
Gogny forces as well as microscopic+macroscopic approaches based on
phenomenological folded Yukawa and Woods-Saxon potentials. As it can
be seen for instance in Refs.\ \cite{AAR.12,AAR.13-epj}, these models
accurately reproduce the inner fission barriers in the actinides.
This is in part due to the fact that the heights of fission barriers
and/or the energy of the fission isomers have been used in their fitting
protocols.

Thus, it is important to understand how these models extrapolate to
the edge on the known region of superheavy nuclei and its vicinity. This
is because the differences in their predictions define the systematic
uncertainties. Fig.\ \ref{FB-SHE} shows the heights of inner fission
barriers of the $Z=112-120$ superheavy nuclei for various relativistic
and non-relativistic models. While providing similar predictions in the
actinides, they do extrapolate in very different ways to the superheavy region.
Their predictions vary significantly and the inner fission barrier heights found
within these models can differ by up to 6 MeV. The substantial differences
in the predictions of the two macroscopic+microscopic (MM) are in particular
surprising. Unfortunately, at present, there are only very few experimental
data available on fission barriers in superheavy elements and they are not
reliable enough to distinguish between theoretical predictions of the
various models (see discussion in Ref.\ \cite{AAR.12}).

  Fig.\ \ref{FB-SHE} also compares the energies of the lowest inner fission barriers
obtained in triaxial RMF+BCS (Ref.\ \cite{AAR.12}) and RHB (present manuscript)
calculations with the CEDF NL3*. Pairing correlations are treated in these two calculations
in a very different way. Monopole pairing with a finite pairing window is used in the
RMF+BCS calculations of Ref. \cite{AAR.12}. Its strength is adjusted  to the
\textquotedblleft empirical\textquotedblright%
\ pairing gaps of Ref.\ \cite{MN.92}.  In the RHB calculations, the separable
form of the finite range Gogny pairing interaction \cite{TMR.09} is used. As discussed
in Sec.\ \ref{theory_details}, this pairing well reproduces physical observables
sensitive to pairing in the actinides. The differences in the calculated inner fission
barriers seen in Fig.\ \ref{FB-SHE} are (i) due to different extrapolation properties
of these two types of pairing on going from actinides to the superheavy region and (ii)
due to the dependence of fission barrier heights on the pairing window used for the
monopole force \cite{KALR.10}. Because of these reasons the inner fission barriers are
found to be roughly 1 MeV higher in the RHB results than in the RMF+BCS calculations
for $N\leq 174$ ($N > 176$). For these neutron numbers, the RHB results come closer
to the mic-mac model predictions 'MM (Kowal)'.  However, the difference between
the TRHB and RMF+BCS results decreases at higher $N$. Note that for the $Z=118$ and
120 nuclei the TRHB results are close to the 'MM (Kowal)' results.

  Of course, it is very difficult to measure the fission barriers. On the
other hand, one can consider also the spontaneous fission half-lives
$\tau_{SF}$ which are directly measurable quantities. Their calculations,
however, represents a real challenge. The values of spontaneous fission
half-lives depend strongly on the underlying theory used to describe the
collective motion, such as the generator coordinate method (GCM) or the
adiabatic time-dependent HFB (ATDHFB) theory (for details see Refs.
\cite{RS.80,SR.16}) and the corresponding collective Hamiltonian, in  particular,
on the inertia parameters. Typical differences between the $\tau_{SF}$ values
calculated with ATDHFB and GCM can reach many orders of magnitude
\cite{RR.14}.

   In addition, the uncertainties (both systematic and statistical) in the
calculated heights of inner fission barriers discussed above will also have
a  profound effect on the calculated spontaneous fission half-lives. For example,
it is well known that the change of fission barrier height by 1 MeV leads to a
change of the calculated spontaneous fission half-lives $\tau_{SF}$ by 6 orders
of magnitude \cite{RR.14}. It is more difficult to quantify the impact of the
change of the topology of the PES on  $\tau_{SF}$, but it is reasonable to expect
that it is substantial.

  As a result, the absolute values of calculated spontaneous fission
half-lives $\tau_{SF}$ cannot be used with confidence since they have
extremely large theoretical uncertainties spanning many orders of magnitude.
However, it is frequently argued that isotopic and/or isotonic trends in
the description of spontaneous half-lives are expected to be reproduced
with much higher accuracy \cite{RR.14}. However, such arguments are
usually based on a single functional. On the contrary, the current
analysis based on a set of the state-of-the art CEDFs as well as
the comparison with other models shown in Fig.\ \ref{FB-SHE} indicates substantial
theoretical uncertainties in isotopic and isotonic trends for the inner
fission barriers, even for the functionals which are benchmarked in the
actinides. In addition, these uncertainties have a ``chaotic'' component
which randomly changes from nucleus to nucleus. These uncertainties will
definitely affect the calculated spontaneous fission half-lives
by many orders of magnitude. This fact is important not only for
our understanding of SHEs but also for fission recycling in neutron
star mergers \cite{GBJ.11}. The later process will be definitely
affected by the increased (as compared with the actinides) uncertainties
of the inner fission barrier heights seen in neutron-rich nuclei (see Fig.\
\ref{fission_spread}).

\section{Conclusions}
\label{concl}

  Theoretical uncertainties in the predictions  of inner fission
barrier heights in SHEs have been investigated for the first time
in a systematic way for covariant energy density functionals.  The
analysis is based on the state-of-the-art functionals NL3*, DD-ME2,
DD-ME$\delta$, DD-PC1, and PC-PK1 which represent major classes of
CEDFs with different basic model assumptions and fitting protocols.
These functionals have been used earlier in the assessment of
theoretical uncertainties in the description of various ground
state observables in Refs.\
\cite{AARR.13,AARR.14,AANR.15,AAR.16,AAR.16,AA.16-prep}. The following
results have been obtained:
\begin{itemize}
\item
  Systematic theoretical uncertainties in the predictions of inner
fission barriers and their propagation towards unknown regions of
higher $Z$ values and of more neutron-rich nuclei have been quantified.
These uncertainties are substantial in SHEs. Statistical uncertainties
are smaller than systematic ones. It is clear that the differences
in the basic model assumptions such as a range of the interaction and
the form of the density dependence together with the different fitting
protocols based  only on nuclear matter and bulk properties data lead
to these uncertainties.

\item
  Systematic theoretical uncertainties in the inner fission barrier
heights do not form a smooth function of proton and neutron numbers;
there is always a random component in their behavior. This is a
consequence of the fact that fission barrier height is the difference
of the energies between the ground state and saddle point. Any
differences in the predictions of their energies, which are not acting
coherently as a function of proton and neutron numbers, will lead to
this random component.

\item
 Benchmarking of the functionals to the experimental data on fission barriers
in the actinides allows to reduce the theoretical uncertainties for the inner
fission barriers of unknown SHEs. However, even then they increase on moving
away from the region where benchmarking has been performed. This feature is
seen not only for different CEDFs but also for different
classes of the models such as microscopic+macroscopic and non-relativistic DFTs.
The resulting uncertainties in the heights of inner fission barriers will
result in uncertainties of many orders of magnitude for spontaneous fission
half-lives. The increased theoretical uncertainties in the fission barriers of
neutron-rich SHEs could have a substantial impact on fission recycling modeling
in r-process simulations of neutron-star mergers.

\item
  Comparing different functionals one can see that the results
(including the topology of the PES) obtained with DD-ME$\delta$  differ
substantially from the results of other functionals. The heights of the
inner fission barriers obtained with this functional are significantly
lower than the experimental estimates in the $Z=112-116$ nuclei and the
values calculated in all other models. In addition, this
functional does not lead to octupole deformation in those actinides which are
known to be octupole deformed \cite{AAR.16}. Thus, this functional
is not recommended for future investigations in the actinides and
superheavy nuclei in spite of the fact that it  provides a good
description of masses and other ground state observables in the
$Z\leq 82$ nuclei \cite{AARR.14}.

\end{itemize}

  The analysis of the description of fission barrier heights is
frequently performed in terms of the parameters which are related
to bulk properties (see, for example, the discussion in Ref.\
\cite{BKRRSW.15}). However, this is only part of the physics which
affects the heights of fission barriers. Indeed, it is well known
that in actinides the lowering of the inner and outer fission
barriers due to triaxial and octupole deformations is caused by
relevant changes in the single-particle density which affect the shell
correction energy \cite{MSI.09,AAR.12}. Substantial differences in the
predictions of the ground states deformations by the state-of-the-art
CEDFs along the $Z=120$ and $N=184$ lines (see Ref.\ \cite{AANR.15})
are also caused by the differences in the underlying single-particle
structure. The differences among the models in the single-particle
structure of superheavy nuclei are substantially higher than in the
region of known nuclei \cite{BRRMG.99,AANR.15}. It is clear that
this is one of the major contributors to the systematic theoretical
uncertainties in the description of inner fission barriers. A further
improvement in the description of the single-particle energies within
DFT is needed in order to reduce the systematic theoretical uncertainties
in the description of fission barriers.

\begin{acknowledgments}
  This material is based upon work supported by the Department of
Energy National Nuclear Security Administration under Award Number
DE-NA0002925, by the U.S. Department of Energy, Office of Science,
Office of Nuclear Physics under Award Number DE-SC0013037 and by
the DFG cluster of excellence \textquotedblleft Origin and Structure
of the Universe\textquotedblright\
(www.universe-cluster.de).
\end{acknowledgments}

\bibliography{references16}

\end{document}